%% file: DESY-06-166.tex
\documentclass[zpreprint,zbstnp]{zeus_paper}
\usepackage[english]{babel}

\input zeus_def.tex

\def\citeCTD{{\cite{%
nim:a279:290,*npps:b32:181,*nim:a338:254%
}}\xspace}
\def\citeCAL{{\cite{%
nim:a309:77,*nim:a309:101,*nim:a321:356,*nim:a336:23%
}}\xspace}

\begin{document}
\prepnum{{DESY--06--166}}
\title{Measurement of open beauty production at HERA in the $D^*\mu$ final state}
      
\author{ZEUS Collaboration} 

\date{September 2006} 

\abstract{The production of beauty quarks with a $D^{*\pm}$ and a muon
  in the final state has been measured with the ZEUS detector at HERA
  using an integrated luminosity of 114 pb$^{-1}$.  Low
  transverse-momentum thresholds for the muon and $D^*$ meson allow a
  measurement of beauty production closer to the production threshold
  than previous measurements.  The beauty signal was extracted using
  the charge correlations and angular distributions of the muon with
  respect to the $D^*$ meson.  Cross sections for photoproduction and
  deep inelastic scattering are somewhat higher than, but compatible
  with, next-to-leading-order QCD predictions, and compatible with
  other measurements.}

\makezeustitle

\def\3{\ss}
\pagenumbering{Roman}
\begin{center}
{                      \Large  The ZEUS Collaboration              }
\end{center}

  S.~Chekanov$^{   1}$,                                                                            
  M.~Derrick,                                                                                      
  S.~Magill,                                                                                       
  S.~Miglioranzi$^{   2}$,                                                                         
  B.~Musgrave,                                                                                     
  D.~Nicholass$^{   2}$,                                                                           
  \mbox{J.~Repond},                                                                                
  R.~Yoshida\\                                                                                     
 {\it Argonne National Laboratory, Argonne, Illinois 60439-4815}, USA~$^{n}$                       
\par \filbreak                                                                                     
  M.C.K.~Mattingly \\                                                                              
 {\it Andrews University, Berrien Springs, Michigan 49104-0380}, USA                               
\par \filbreak                                                                                     
  N.~Pavel~$^{\dagger}$, A.G.~Yag\"ues Molina \\                                                   
  {\it Institut f\"ur Physik der Humboldt-Universit\"at zu Berlin,                                 
           Berlin, Germany}
% ...
\newpage                                                                                           
$^{\    1}$ supported by DESY, Germany \\                                                          
$^{\    2}$ also affiliated with University College London, UK \\                                  
% ...
$^{\dagger}$ deceased \\                                                                           
%
\newpage   
                                                           %
                                                           %
\begin{tabular}[h]{rp{14cm}}                                                                       
$^{a}$ &  supported by the Natural Sciences and Engineering Research Council of Canada (NSERC) \\  
$^{b}$ &  supported by the German Federal Ministry for Education and Research (BMBF), under        
          contract numbers HZ1GUA 2, HZ1GUB 0, HZ1PDA 5, HZ1VFA 5\\                                
$^{c}$ &  supported in part by the MINERVA Gesellschaft f\"ur Forschung GmbH, the Israel Science   
          Foundation (grant no. 293/02-11.2) and the U.S.-Israel Binational Science Foundation \\  
$^{d}$ &  supported by the German-Israeli Foundation and the Israel Science Foundation\\           
$^{e}$ &  supported by the Italian National Institute for Nuclear Physics (INFN) \\                
$^{f}$ &  supported by the Japanese Ministry of Education, Culture, Sports, Science and Technology 
          (MEXT) and its grants for Scientific Research\\                                          
$^{g}$ &  supported by the Korean Ministry of Education and Korea Science and Engineering          
          Foundation\\                                                                             
$^{h}$ &  supported by the Netherlands Foundation for Research on Matter (FOM)\\                   
$^{i}$ &  supported by the Polish State Committee for Scientific Research, grant no.               
          620/E-77/SPB/DESY/P-03/DZ 117/2003-2005 and grant no. 1P03B07427/2004-2006\\             
$^{j}$ &  partially supported by the German Federal Ministry for Education and Research (BMBF)\\   
$^{k}$ &  supported by RF Presidential grant N 1685.2003.2 for the leading scientific schools and  
          by the Russian Ministry of Education and Science through its grant for Scientific        
          Research on High Energy Physics\\                                                        
$^{l}$ &  supported by the Spanish Ministry of Education and Science through funds provided by     
          CICYT\\                                                                                  
$^{m}$ &  supported by the Particle Physics and Astronomy Research Council, UK\\                   
$^{n}$ &  supported by the US Department of Energy\\                                               
$^{o}$ &  supported by the US National Science Foundation\\                                        
$^{p}$ &  supported by the Polish Ministry of Science and Higher Education\\                       
$^{q}$ &  supported by FNRS and its associated funds (IISN and FRIA) and by an Inter-University    
          Attraction Poles Programme subsidised by the Belgian Federal Science Policy Office\\     
$^{r}$ &  supported by the Malaysian Ministry of Science, Technology and                           
Innovation/Akademi Sains Malaysia grant SAGA 66-02-03-0048\\                                       
\end{tabular}                                                                                      

\newpage
\pagenumbering{arabic} 
\pagestyle{plain}

\section{Introduction}
\label{sec-int}

The production of beauty quarks in $ep$ collisions at HERA is a
stringent test for perturbative Quantum Chromodynamics (QCD) since the
large $b$-quark mass ($m_b \sim 5 \gev$) provides a hard scale that
should ensure reliable predictions.  For $b$-quark transverse momenta
comparable to the $b$-quark mass, next-to-leading-order (NLO) QCD
calculations in which the $b$ quarks are generated dynamically are
expected to provide accurate predictions
\cite{np:b412:225,*np:b454:3-24,*frixione3,
  np:b452:109,*pl:b353:535,*pr:d57:2806}.
 
The beauty-production cross section has been measured in $p\bar{p}$
collisions at the S$p\pbar$S
\cite{beautyUA10,*pl:b256:121-128,*zfp:c61:41-52} and Tevatron
colliders
\cite{beautyCDF1,*beautyCDF2,*beautyCDF3,*beautyCDF4,*beautyCDF5,
  *beautyCDF6_pap,*beautyD00,*beautyD01,*beautyD02,*beautyD03}, in
$\gamma \gamma$ interactions at LEP \cite{beautyLEP0,*beautyLEP1}, in
fixed-target $\pi N$ \cite{WA78,*E706} and $pN$
\cite{E771,*E789,*HERAB} experiments, and in $ep$ collisions at HERA
\cite{pl:b467:156,*epj:c18:625,epj:c40:349,*epj:c41:453,*Aktas:2005iw,*H1phjets,pr:d70:012008,pl:b599:173,pl:b621:56-71}.
While most results, including recent results from the Tevatron, are in
agreement with QCD predictions, some, in particular those from LEP,
show large discrepancies.

This paper reports a measurement of beauty production via the reaction
$e p \to e b \bar{b} X \to e D^* \mu X^\prime$ using the ZEUS detector
at HERA.  This reaction offers the advantage of providing a data
sample enriched in $b$ quarks and with strongly suppressed backgrounds
from other processes, which allows low-$p_T$ threshold cuts to be
applied.  This analysis therefore yields a measurement of beauty
production closer to the production threshold than previous HERA
measurements based on leptons and/or jets with high transverse
momentum
\cite{pl:b467:156,*epj:c18:625,epj:c40:349,*epj:c41:453,*Aktas:2005iw,*H1phjets,pr:d70:012008,pl:b599:173}.
A similar measurement has been performed by the H1 collaboration
\cite{pl:b621:56-71}.

Of particular interest are events in which the muon and $D^*$
originate from the same parent $B$ meson (Fig.~\ref{fig_topol}a),
e.g. $B^0 \to D^{*-}\mu^+\nu_\mu$. These yield unlike-sign $D^*$-muon
pairs produced in the same detector hemisphere. Due to the partial reconstruction
(e.g. missing neutrino) the invariant mass is constrained to lie
below the $B$-meson mass. Another important contribution arises from
charm-pair production, where one charm quark fragments into a $D^*$
and the other decays into a muon (Fig.~\ref{fig_topol}b). This again
yields unlike-sign $D^*$-muon pairs, but with the $D^*$ and the muon
produced predominantly in opposite hemispheres. In addition,
beauty-pair production in which the $D^*$ and muon originate from
different beauty quarks can yield both like- and unlike-sign
$D^*$-muon combinations, depending on whether the muon is from the
decay of the primary beauty quark (Fig.~\ref{fig_topol}c), or from a
secondary charm quark (Fig.~\ref{fig_topol}d), and whether
$B^0$-$\bar{B^0}$ mixing has occured.

Background contributions to both like- and unlike-sign combinations
include events with either fake $D^*$ mesons, originating from
combinatorial background, or fake muons.  In this analysis, the signal
is extracted from the unlike-sign sample, while the like-sign sample
is used as a cross check.

Cross sections are extracted separately for the photoproduction
($\gamma p$, photon virtuality $Q^2 \lesssim 1$ GeV$^2$), and deep
inelastic scattering (DIS, $Q^2 \gtrsim 1$ GeV$^2$) regimes, as well
as for the entire range in $Q^2$, which includes the kinematic region
in which these two event classes cannot easily be distinguished.

\section{Experimental set-up}
\label{sec-exp}
The data sample used in this analysis corresponds to an integrated
luminosity ${\cal L}=114.1 \pm 2.3 ~\rm{pb}^{-1}$, collected by the
ZEUS detector in the years 1996-2000.  During the 1996-97 data taking,
HERA provided collisions between an electron\footnote{Electrons and
  positrons are not distinguished in this paper and are both referred
  to as electrons.}  beam of $E_e=27.5 \gev$ and a proton beam of
$E_p=820\gev$, corresponding to a centre-of-mass energy $\sqrt
s=300\gev$ (${{\cal L}_{300}}=38.0\pm 0.6~ \rm{pb}^{-1}$). In the
years 1998-2000, the proton-beam energy was $E_p=920\gev$,
corresponding to $\sqrt s=318\gev$ (${{\cal L}_{318}}=76.1\pm
1.7~\rm{pb}^{-1}$).

A detailed description of the ZEUS detector can be found
elsewhere~\cite{zeus:1993:bluebook}. A brief outline of the components
that are most relevant for this analysis is given below.

\Zctddesc\ZcoosysfnBeta

The high-resolution uranium-scintillator calorimeter (CAL)~\citeCAL
consists of three parts: the forward (FCAL), the barrel (BCAL) and the
rear (RCAL) calorimeters.  Each part is subdivided transversely into
towers and longitudinally into one electromagnetic section and either
one (in RCAL) or two (in BCAL and FCAL) hadronic sections.  The
smallest subdivision of the calorimeter is called a cell.  The CAL
energy resolutions, as measured under test-beam conditions, are
$\sigma(E)/E=0.18/\sqrt{E}$ for electrons and
$\sigma(E)/E=0.35/\sqrt{E}$ for hadrons, with $E$ in $\Gev$.

The position of electrons scattered at small angles to the electron
beam direction was measured using the small-angle rear tracking
detector (SRTD) \cite{nim:a401:63,epj:c21:443}.  The SRTD is attached
to the front face of the RCAL and consists of two planes of
scintillator strips, arranged orthogonally. The strips are 1 cm wide
and 0.5 cm thick.

The muon system consists of rear, barrel (R/BMUON) \cite{nim:a333:342}
and forward (FMUON) \cite{zeus:1993:bluebook} tracking detectors.  The
B/RMUON consists of limited-streamer tube chambers placed behind the
BCAL (RCAL), inside and outside a magnetized iron yoke surrounding the
CAL.  These chambers cover polar angles from $34^{\rm o}$ to $135^{\rm
  o}$ and from $135^{\rm o}$ to $171^{\rm o}$, respectively.

The luminosity was measured using the bremsstrahlung process $ep \to e
p \gamma$.  The resulting small-angle energetic photons were measured
by the luminosity
monitor~\cite{Desy-92-066,*zfp:c63:391,*acpp:b32:2025}, a
lead-scintillator calorimeter placed in the HERA tunnel at $Z = -107$
m.

\section{Data Selection}
\label{data_sel}
The data were selected online via a three-level trigger system through
a combination of three different trigger chains:
\begin{itemize}
\item a muon reaching the inner B/RMUON chambers and matched to a
  minimum ionizing energy deposit in the CAL or any muon reaching the
  outer B/RMUON chambers or
\item a $D^*$ candidate \cite{np:b729:492-525} or
\item a scattered-electron candidate in the CAL \cite{pl:b599:173}.
  In part of the data taking, the cuts on the electron candidate were
  relaxed if a muon in the inner B/RMUON chambers was detected.
\end{itemize}
Due to this redundancy, the trigger efficiency for beauty events was
high, $94 \pm 3$\% for the inclusive study, and $98 \pm 2$\% for the
DIS selection.

Muons were reconstructed offline using the following procedure: a
track was found in the inner B/RMUON chambers, then a match in
position and angle to a CTD track was required.  In the bottom region
of the detector, where there are no inner chambers, the outer chambers
were used instead. If a match was found to both inner and outer
chambers, a momentum-matching criterion was added.

The angular coverage of the B/RMUON and of the track requirements in
the CTD restrict the muon acceptance to the pseudorapidity region
\begin{equation}
\label{mu:eta} 
-1.75<\eta^{\mu}<1.3 \ .
\end{equation}
A cut on the muon transverse momentum 
\begin{equation}
\label{mu:pt} 
p_{T}^{\mu}>1.4\gev
\end{equation}
was applied, reflecting the requirement that the muon reaches the
inner muon chambers in the barrel region.  In order to have uniform
kinematic acceptance, this cut was also applied in the rear region.

$D^*$ candidates were reconstructed in the $D^{*+} \to D^0(\to K^-
\pi^+)\pi_s^+$ decay channel (+c.c.)  making use of the $\Delta M$
($\equiv M(K\pi\pi_s)-M(K\pi)$) technique described in previous
publications \cite{np:b729:492-525} with the following cuts:
\begin{eqnarray}
\label{e:dscuts}
 D^0~\textrm{mass} && 1.81 < M(K\pi) < 1.92\gev; \nonumber \\
 D^*-D^0~\textrm{mass difference} && 0.1435 <
               \Delta M < 0.1475\gev; \nonumber \\
 D^*~\textrm{transverse momentum} &&  p_T^{D^*} > 1.9\gev; \\
 D^*~\textrm{pseudorapidity} &&  \vert\eta^{D^*}\vert<1.5; \nonumber \\
 K,\pi~\textrm{transverse momentum} && p_T^{K,\pi} > 0.5\gev; \nonumber \\
 \textrm{slow pion} && p_T^{\pi_s} > 0.125\gev. \nonumber
\end{eqnarray}
To allow the background to the $D^*$ signal to be determined, $D^0$
candidates with wrong-charge combinations, in which both tracks
forming the $D^0$ candidates have the same charge and the third track
has the opposite charge, were also retained.

The hadronic system was reconstructed from the calorimeter information
and the reconstructed vertex.  A four-momentum
$(p_X^i,p_Y^i,p_Z^i,E^i)$ was assigned to each calorimeter cell.
Global hadronic variables were reconstructed by summing over these
cells.  In the case of identified DIS events (see below), the
scattered electron candidates were excluded from this sum.  The
inelasticity $y$ was reconstructed from the Jacquet-Blondel estimator
$y_{\rm JB}=(E-P_Z)/2E_e$ \cite{proc:epfacility:1979:391}, where
\mbox{$E-P_Z=\sum_i (E^i-p_Z^i)$} and the sum runs over all cells.  In
the case of DIS events, the alternative value
$y_e=1-\frac{E_e^\prime}{2E_e} (1-\cos \theta_{e})$ as well as the
photon virtuality $Q^2$ were obtained from the energy $E_e^\prime$ and
scattering angle $\theta_e$ of the final-state electron candidate
\cite{pl:b599:173}.  A sample of events with one muon and one $D^*$
candidate was selected by requiring:
\begin{itemize}
\item $\ge 1$ muon in the muon chamber regions defined by
  Eq.~(\ref{mu:eta}) and Eq.~(\ref{mu:pt});
\item $\ge 1$ $D^*$ candidate in the $D^*$ acceptance region defined
  by Eq.~(\ref{e:dscuts});
\item the muon candidate track is not one of the three $D^*$ candidate
  tracks, eliminating backgrounds from semileptonic $D^0$ decays;
\item the $D^*\mu$ ~system carries a significant fraction of the total
  transverse energy of the event, $p_T^{D^*\mu}/E_T>0.14$, where $E_T$
  is the transverse energy measured by the CAL outside a cone of
  10$^\circ$ around the proton beam direction to exclude the proton
  remnant, and $p_T^{D^*\mu}$ is the transverse momentum of the
  $D^*\mu$ system, reducing combinatorial $D^*$ background;
\item a reconstructed vertex compatible with the nominal interaction
  point, suppressing non-$ep$ background.
\end{itemize}

After this selection, a sample of 232 events remained.  The resulting
$\Delta M$ distributions for the like and unlike $D^*\mu$ charge
combinations, before the $\Delta M$ cut, are shown in
Figs.~\ref{fig1}a and \ref{fig1}b.

A subsample of photoproduction events was selected by requiring:
\begin{itemize}
\item no scattered-electron candidate found in the CAL;
\item $E-P_Z < 34 \gev$;
\end{itemize}
retaining 81\% of the inclusive sample.  After the unfolding of the
detector response, these cuts correspond to an effective cut
$Q^2<1\gev^2$ and $0.05<y<0.85$. The lower limit on $y$ arises from
the interplay between the $b$-quark mass and the acceptance in
rapidity.

Alternatively, a clean DIS sample was obtained by applying the
following additional conditions \cite{pl:b599:173}:
\begin{itemize}
\item a reconstructed electron with energy $E_e^\prime >$10 GeV;
\item $Q^2 > 2$ GeV$^2$;
\item inelasticity $y_{\rm{JB}}>0.05$ and $y_{e} < 0.7$;
\item $40< E-P_Z < 60 \gev$;
\item the electron hits the rear calorimeter outside a rectangle of 
      $|X|<13$ cm and $|Y|<7$ cm. 
\end{itemize}
These cuts correspond to an effective cut $Q^2>2\gev^2$ and
$0.05<y<0.7$.  For this sample, which contains less combinatorial
background, the $D^*$ cuts were relaxed to
\begin{eqnarray}
\label{e:dsDIS}
p_T^{D^*} &>& 1.5\gev; \nonumber \\ 
p_T^{K,\pi} &>& 0.4\gev; \\
p_T^{\pi_s} &>& 0.12\gev; \nonumber
\end{eqnarray}
and the cut on $p_T^{D^*\mu}/E_T$ was dropped. All other cuts on the
$D^*$ and the muon remained unchanged. A sample of 44 events was
obtained.  The resulting $\Delta M$ distributions for the like and
unlike $D^*\mu$ charge combinations are shown in Figs.~\ref{fig1}c and
\ref{fig1}d.

\section{Backgrounds and event simulation}

Several contributions to the selected data sample were evaluated:
\begin{itemize}
\item the signal from beauty decays;
\item the background from fake $D^*$ combinations;
\item the $D^*\mu$ background from charm decays;
\item the background from fake or non-prompt muons with a real $D^*$ from charm.
\end{itemize}

For the signal from beauty and charm production, Monte Carlo (MC)
simulations were performed using the {\sc Pythia} \cite{cpc:82:74},
{\sc Rapgap} \cite{cpc:86:147} and {\sc Herwig} \cite{cpc:67:465}
generators.  These simulations are based on leading-order matrix
elements complemented by parton showers to obtain higher-order
topologies.  The direct photon-gluon fusion process ($\gamma g \to Q
\bar Q$, $Q=b,c$), flavour excitation in the resolved photon and
proton (e.g. $Q g \to Q g$, $\gamma Q \to Q g$), and hadron-like
resolved photon processes ($g g \to Q \bar{Q}$) were included.  Gluon
splitting into heavy flavours ($g\to Q\bar Q$) in events with only
light quarks in the hard scattering was not included in the
simulations; this contribution is, however, expected to be small
\cite{thesis:longhin:2004}.  For all generated events, the ZEUS
detector response was simulated in detail using a programme based on
GEANT 3.21 \cite{GEANT}.

The number of background events under the $D^*$ mass peaks (fake
$D^*$) was estimated using the wrong-charge $K\pi$ combinations
combining the like- and unlike-sign samples.  This was found to
minimize the bias due to charge correlations
\cite{thesis:longhin:2004}.  Wrong-charge combinations were normalised
to the data outside the $D^*$ peak in the side-band $0.15 < \Delta M <
0.17\gev$, separately for the like-sign and unlike-sign $D^*\mu$
sample, as shown in Fig.~\ref{fig1}.  Dedicated studies
\cite{thesis:longhin:2004} performed by selecting data on the $D^*$
side-band showed that this procedure correctly reproduces shape and
normalisation of the fake-$D^*$ background for the relevant variables
used in the analysis.

Fake muons can be produced by hadron showers leaking from the back of
the calorimeter or by charged hadrons traversing the entire
calorimeter without interaction.  In addition, low-momentum muons can
originate from in-flight decays of pions and kaons. It is also
possible for tracks reconstructed in the central tracker to be wrongly
associated to a signal from a real muon in the muon chambers.  A
dedicated study \cite{thesis:longhin:2004} based on pions from $K^0$
decays, protons from $\Lambda$ decays, and kaons from $\phi$ and $D^*$
decays, showed that the detector simulation reproduced these
backgrounds reasonably well. 
The fake muon probability for the $K^0\to\pi^+\pi^-$ sample is about 0.2\%. 
Most fake muons are associated with fake
$D^*$ candidates, and therefore accounted for in the fake-$D^*$
background estimated directly from the data. Fake muons associated
with a real $D^*$ are included in the charm and beauty MC samples.

Distributions of variables used in the event selection or relevant for
the event kinematics, for the unlike-sign inclusive sample, are
compared to the expectations from these simulations in
Fig.~\ref{fig:controlplots}, separately for the beauty- and
charm-enriched regions defined in Section \ref{sect:signal}. Agreement
with expectations is obtained, 
apart from some possible deviations in the $p_T^{D^*\mu}$ and
$p_T^{D^*\mu}/E_T$ distributions in the beauty-enriched region, which
are accounted for in the systematic uncertainties (Section
\ref{sect:syst}).

\section{Signal extraction} 
\label{sect:signal}

In this section, the signal-extraction procedure is described for the
inclusive sample. The $\gamma p$ subsample and the DIS sample were
treated in an analogous way.

Figures \ref{fig2}a and \ref{fig2}b show the distribution of the
angular difference $\Delta R = \sqrt{\Delta \phi^2 + \Delta \eta^2}$
between the $D^*$ and the muon, where $\phi$ is the azimuthal angle,
for events passing all selections, including the $\Delta M$ cut.  The
distributions are shown separately for like- and unlike-sign $D^*\mu$
events. The expected signal and background distributions, normalised
to the fractions determined later in the analysis and described below,
are also indicated.  For unlike-sign events, the region $\Delta R >
2$, which mainly corresponds to the back-to-back configuration, is
clearly dominated by events from charm. Indications that the simulated
distribution might be more sharply peaked than the data turned out to
have little influence on the determination of the beauty fraction.  In
contrast, the region $\Delta R < 2$ is enriched in beauty events, in
which the $D^*$ and muon originate mainly from decays of the same
parent $B$ hadron.  This is illustrated further in the $D^*\mu$
invariant-mass distribution (Figs.~\ref{fig2}c and \ref{fig2}d) for
events in the beauty-enriched region ($\Delta R < 2$).  A peak with an
upper edge close to 5 GeV, which can be attributed to the partial
reconstruction of the decaying $B$ meson, is clearly visible. A
comparison with the like-sign sample shows that the low-mass edge of
this peak is dominated by background. An invariant-mass cut of $3~{\rm
  GeV} < M(D^*\mu) <5~{\rm GeV}$ was therefore applied to the $\Delta
R < 2$ subsample.

After this additional cut, and after statistical subtraction
of the fake-$D^*$ background, the contributions of charm and
beauty were determined by a two-component fit to the $\Delta R$ or $\Delta \phi$
distributions, shown in Figs.~\ref{fig3}a and \ref{fig3}c.
%
%
The fake-muon background with a real $D^*$ from charm, which is
treated as part of the charm component, is shown separately.  The
small fraction of fake muons from beauty was included in the beauty
component.  The fit result for the fraction of beauty events in the
final inclusive sample shown in Figs.~\ref{fig3}a and \ref{fig3}c,
using the shapes predicted by {\sc Pythia}, is:
\begin{itemize}
\item $f_b = 0.307 \pm 0.064{\rm (stat.)}~{\rm for~the}~\Delta R$ and  
\item $f_b = 0.290 \pm 0.062{\rm (stat.)}~{\rm for~the}~\Delta \phi$ distribution. 
\end{itemize}
The $\Delta R$ result was chosen as the reference, and the $\Delta
\phi$ result used as a systematic check.  With these fitted fractions,
the breakdown into the corresponding number of beauty, charm, and
fake-muon candidates in each subsample is given in Table
\ref{tab_evts}.  In the unlike-sign part, the beauty and charm
contributions are well separated, with only small cross-contaminations.
The normalisation of the beauty and charm contributions in
Fig.~\ref{fig2} already reflects these fitted fractions.  Agreement is
seen, also in the like-sign part, which was not included in the fit.

The results from the application of the same procedure to the $\gamma
p$ subsample are also shown in Table \ref{tab_evts}.  The analogous
results for the DIS sample are shown in Figs.~\ref{fig3}b and
\ref{fig3}d.  The $\Delta \phi$ distribution gives less discrimination
in this case, due to the transverse boost from the exchanged virtual
photon.  Therefore, the $M(D^*\mu)$ distribution was used.  The fitted
beauty fractions in the DIS sample, using the shapes predicted by {\sc
  Rapgap}, are
\begin{itemize}
\item $f_b = 0.55 \pm 0.25\rm{(stat.)}$ for the $\Delta R$ and
\item $f_b = 0.43 \pm 0.30\rm{(stat.)}$ for the $M(D^*\mu)$ distribution. 
\end{itemize}  
Again, the $\Delta R$ result is chosen as the reference, and the other
as a cross check. The breakdown into different event categories is
shown in Table \ref{tab_evts}.  The acceptance corrections for the
cross sections which will be presented in Section \ref{sec:results}
were evaluated using {\sc Pythia} for $Q^2<1\gev ^2$, {\sc Rapgap} for
$Q^2>1\gev ^2$, and {\sc Herwig} as a systematic check.

\section{Theoretical predictions and uncertainties}

For direct comparisons with QCD, next-to-leading-order predictions
were used.  Calculations in which $b$ quarks are treated as massless
particles \cite{bmassless1,*bmassless2,*bmassless3} are not applicable
in this kinematic range, while calculations based on alternative
parton-evolution schemes \cite{bkt1,*bkt2,*jung1,*jung2} do not yet
exist with full NLO implementation. Fixed-order NLO calculations with
massive $b$ quarks should yield accurate predictions.  Different types
of such predictions were evaluated.

The FMNR program \cite{np:b412:225} evaluates cross sections for
next-to-leading-order beauty production in $\gamma p$ collisions in
the fixed-order massive approach, for both point-like and hadron-like
photon coupling to the heavy quarks.  The parton-density functions
used were CTEQ5M \cite{epj:c12:375} for the proton and GRV-G-HO
\cite{pr:d46:1973-1979} for the photon. The renormalisation and
factorisation scales $\mu$ were chosen to be equal and parametrised by
$\mu_0 = \sqrt{p_{T}^2+m_b^2}$, where $p_T^2$ is the average of the
squared transverse momentum of the two emerging $b$ quarks and
$m_b=4.75$ GeV. An estimate of the theoretical uncertainty was
obtained by simultaneously varying $4.5 < m_b < 5.0$ GeV and $\mu_0/2
< \mu < 2\mu_0$ such that the uncertainty was maximised. Typical
uncertainties resulting from this procedure (e.g. for the $b\bar b$
total cross section) are +40\% and -25\%.  Variations of the parton
densities led to uncertainties which were much smaller than the
uncertainties related to mass and scale variations. They were
therefore neglected.

Predictions at the level of visible $D^*\mu$ final states are needed
in addition to those at parton level.  The FMNR program provides a
framework to fragment $b$ quarks into $B$ hadrons, and simulate the
decay of these hadrons by interfacing them to appropriately chosen
decay spectra. However, decays to complex final states, such as a
$D^*$ and $\mu$ from the same $B$ hadron with cuts on both particles,
cannot be easily implemented in this scheme.  A straightforward
interface of the parton-level events produced by FMNR to MC-like
fragmentation and decay chains is also impracticable, since these
events have positive and negative weights spanning more than 8 orders
of magnitude, making such an approach extremely inefficient.

These difficulties were overcome in a two-step process.  In the first
step, two or more FMNR parton-level events with large positive and
negative weights and similar topology were combined into events with
much smaller weights by averaging the parton momentum vectors
\cite{LizDIS}. Events were considered to have similar topology if the
differences in transverse momentum, rapidity and azimuthal angle of
the $b$ quarks were less than user cut values that reflect the
detector resolution.  Furthermore, events with small weights were
sampled with a probability proportional to their weight.  In this way,
the weight range was reduced to about two orders of magnitude.  It was
explicitly checked that this procedure preserves the NLO accuracy for
the relevant cross sections at parton level (e.g. $b$ quark $p_T$ and
angular distributions).

In the second step, these parton-level events were interfaced to the
{\sc Pythia/Jetset} \cite{hep-ph-0108264} fragmentation and decay
chain, making use of the full decay tables and decay kinematics
implemented in {\sc Pythia 6.2}.  Therefore, non-dominant complex
decays, such as $B \to D^* D$ followed by $D \to \mu X$, or muons
produced through intermediate $J/\psi$ or $\tau$ states, were
automatically included.  The initial-state partons were allowed to
have intrinsic $k_T$ (typically $\sim 300$ MeV) as implemented in {\sc
  Pythia}. This has a negligible effect on the resulting cross
sections ($\sim 1$\%).  Parton showering was not included in order to
avoid double counting of higher-order contributions\footnote{The {\sc
    MC@NLO} approach \cite{jhep:06:029,*jhep:08:007}, which allows the
  combination of NLO matrix elements with parton showers, is not yet
  available for $ep$ interactions.  }.

Fragmentation of $b$ quarks close to production threshold is
non-trivial. The details of the threshold treatment were found to be
much more important than the choice of a particular fragmentation
function. The Peterson formula \cite{pr:d27:105} with $\epsilon =
0.0035$ was used.  Three approaches were considered:
\begin{itemize}
\item independent fragmentation as implemented in {\sc Pythia}
  \cite{hep-ph-0108264}.  The use of this quite old model was
  motivated by the fact that FMNR does not provide colour connections
  on an event-by-event basis;
\item fragmentation in the Lund string model
  \cite{zfp:c20:317,*np:b238:492,*cpc:39:347}, again as implemented in
       {\sc Pythia}.  For this purpose, reasonable colour connections
       were assigned to each event;
\item the independent fragmentation scheme provided within the FMNR
  framework, rescaling the $B$-hadron momentum to a fraction of the
  $b$-quark momentum according to the Peterson formula, which is a
  somewhat crude approximation at threshold.
  
\end{itemize}
The second option was used for all central predictions. The first
option was used to obtain the lower systematic error (typically -5\%).
The third option could not be applied directly, since it does not
provide cross-section predictions for correlated final states from the
same $b$ quark, as needed here. Instead, it was applied to a cross
section in which the final-state correlations originate from different
$b$ quarks only, which is more easily calculable in this scheme. The
results were used to evaluate a generic upper systematic error of
+15\% on the fragmentation procedure close to $b$ production
threshold. The effect of a variation of the Peterson parameter
$\epsilon$ in the range 0.0023 to 0.0045 was found to yield
uncertainties that were much smaller than the uncertainties due to the
different fragmentation procedures. It was therefore neglected.

The branching fractions were corrected to correspond to those obtained
from the Particle Data Group \cite{pl:b592:1}, as listed in Table
\ref{tab:branchings}.  Branching fraction uncertainties resulted in
uncertainties on the $D^*\mu$ cross section of typically $\pm 12\%$.

In principle, FMNR predictions are only valid for the photoproduction
regime. The Weizs\"acker-Williams approximation with an effective
$Q^2_{\rm max}$ cutoff of 25 GeV$^2$ ($\sim m_b^2$)
\cite{zfp:88:612,*pr:45:729,*pl:b319:339-345} was used to include the
$\sim$15\% DIS contribution to the combined cross section.

Alternatively, the DIS part can be calculated using the NLO
predictions from HVQDIS \cite{np:b452:109,*pl:b353:535,*pr:d57:2806}.
Only point-like contributions are included in this prediction.  The
parton density function used was CTEQ5F4 \cite{epj:c12:375}.  The
renormalisation and factorisation scales $\mu$ were chosen to be equal
and parametrised by $\mu_0 = \sqrt{Q^2+m_b^2}$.  Mass and scales were
varied as for FMNR.  A scheme for the calculation of visible cross
sections for correlated final states, corresponding to the
FMNR$\otimes${\sc Pythia} interface described above, was not
available. Therefore, cross-section comparisons in DIS are limited to
the parton level.

\section{Systematic uncertainties}
\label{sect:syst}

The main experimental uncertainties are described below, in order of
importance.  Numbers in parentheses are quoted for the inclusive
selection.  Uncertainties for the $\gamma p$ results are also quoted
when they differ significantly from the inclusive results.  For the
DIS sample, the statistics were often too small to derive meaningful
systematic errors. The errors from the inclusive sample were used
instead.

\begin{itemize}
\item {\bf Simulation of $\bf {p_T^{D^*\mu}}$}.  The largest error
  arises from the observation that the muon and $D^*$ $p_T$ spectra in
  the $b$ signal region of the data ($\Delta R(D^*\mu) < 2$,
  $3<M(D^*\mu)<5$ GeV) appear to be somewhat softer than predicted by
  the Monte Carlo simulations (Fig.~\ref{fig:controlplots}).  The
  differences are concentrated at small values of $p_T^{D^*\mu}/E_T$.
  Since the corresponding spectra are well reproduced in the charm
  region with larger statistics, this cannot be attributed to problems
  with the muon or $D^*$ reconstruction. There are several ways to
  interpret these differences:
\begin{itemize}
\item[a)] they are statistical fluctuations. This assumption leads to
  the central result reported;
\item[b)] the signal distribution is significantly softer than
  predicted by QCD. Due to the rising efficiency as a function of
  $p_T^b$, this would change the efficiency calculation for the
  measurement of the visible cross section. To evaluate this
  possibility, the MC $p_T^{D^*\mu}$ (true level) distribution in the
  signal region was reweighted to be compatible at the 1 $\sigma$
  level with the measured $p_T^{D^*\mu}$ spectrum of the inclusive
  sample (Fig.~\ref{fig:controlplots}) (+14\%);
\item[c)] there is an additional unknown background contribution at
  low $p_T$, which occurs only in the beauty-enriched region. There is
  no indication that this is the case.  Nevertheless, to account for
  this possibility, the $p_T^{D^*\mu}/E_T$ cut was tightened from 0.14
  to 0.2, which removes most of the differences
  (Fig.~\ref{fig:controlplots}) (-33\% for inclusive, -18\% for
  $\gamma p$ selection).
\end{itemize}

\item {\bf Branching fractions}. The beauty-enriched region, which
  dominates the fit result, is mainly populated by events in which the
  $D^*$ and $\mu$ originate from the same $b$. The rate of these
  events depends on different branching fractions from those relevant
  to the charm-enriched region, in which the $D^*$ and $\mu$ originate
  from different $b$ quarks. A variation of these branching fractions,
  within the uncertainties quoted in Table \ref{tab:branchings},
  therefore affects the shape of the beauty contribution and the
  fitted beauty fraction ($\pm 8$\%).

\item {\bf Fragmentation and parton showering}. The {\sc Herwig} MC
  uses a different fragmentation model from that of {\sc Pythia} and
  {\sc Rapgap}.  It also yields different $b\bar{b}$ correlations from
  direct/resolved contributions and parton showering. This leads to
  differences in the acceptance, and in the fitted beauty fraction
  (+5/-8\%).

\item {\bf Signal-extraction procedure}.  In addition to statistical
  fluctuations, different ways to fit the data can yield systematic
  differences due to binning effects and different systematics for
  different variables, e.g. imperfections in the shape of the MC
  distributions. To check the error from this effect, the cross
  sections were evaluated using different procedures: fits to $\Delta
  R(D^*\mu)$, $\Delta \phi(D^*\mu)$, $M(D^*\mu)$, and simple event
  counting.  In all cases the differences were well within the quoted
  errors. To avoid double counting of statistical and systematic
  errors, these were used as cross checks only.

\item {\bf Uncertainty on the estimation of the muon chamber
  efficiency}.  Corrections to the MC muon chamber reconstruction
  efficiency were obtained from independent data samples and varied
  within their uncertainties ($\pm 5\%$).

\item {\bf Fake muon background}. The background from fake muons has
  been extensively studied \cite{thesis:longhin:2004} and is further
  constrained by the like-sign distribution of Fig.~\ref{fig2}b, which
  is dominated by this background.  Accordingly, it was varied by a
  factor 1.5 (-4\%).

\item {\bf Luminosity measurement}. The uncertainty associated with
  the luminosity measurement for the 1996-00 data taking periods used
  for this analysis was included ($\pm$2\%).

\item {\bf Tracking}.  All tracking-based cuts ($p_T$ and mass cuts)
  were varied by their respective uncertainties.  To avoid double
  counting of statistical uncertainties, the $D^*$-related systematics
  were taken from previous ZEUS DIS \cite{pr:d69:012004} and $\gamma
  p$ \cite{pl:b590:143} analyses employing similar cuts but with
  larger event samples.  The cut on $p_T^\mu$ was varied by $\pm$40
  MeV.  This yielded a combined error of +6\% and -4\%.

\item {\bf Trigger acceptance}. The error on the trigger acceptance
  was evaluated by comparing the efficiencies of the different trigger
  chains in the data with each other and with the MC ($\pm$3\%).
\item {\bf $B^0$-$\bar{B^0}$ mixing}. The possible systematic effect due
  to the variation of the mixing rate was found to be negligible.
\end{itemize}
The total systematic uncertainty was obtained by adding the above
contributions in quadrature.

\section{Results}     
\label{sec:results}

To present results from the combined data sets, the measurements from
the 1996-97 run at $\sqrt{s} = 300 \gev$ have been corrected using the
predicted cross section ratio \cite{np:b412:225} of 1.06, to
correspond to the higher center-of-mass energy of 318~GeV.  All cross
sections are therefore quoted for $\sqrt{s}$ = 318~GeV.

\subsection{Visible cross sections}

The first step is the extraction of visible cross sections for the
$D^*\mu$ final state from beauty.  The acceptance for the $D^* \to
D^0\pi_s \to (K\pi)\pi_s$ decay chain was unfolded using a branching
fraction of $2.57\pm 0.06$\% \cite{pl:b592:1}.  The effective $b$
branching fractions used in the different MC generators were corrected
to those listed in Table \ref{tab:branchings} in order to account for
their influence on the overall acceptance, and on the shape of the
predicted beauty contributions.

The measured beauty fraction in the inclusive sample, corrected for
detector acceptance and branching fractions, was used to obtain the
cross section for the process $ep \to e b \bar{b} X \to e D^{*\pm}\mu
X$ in the visible kinematic range $p_T^{D^*} > 1.9$ GeV, $-1.5 <
\eta^{D^*} < 1.5$, $p_T^{\mu} > 1.4$ GeV and $-1.75 < \eta^{\mu} <
1.3$ as:
\begin{equation}
\label{sig:tot}
\sigma_{\rm vis}(ep \to e b \bar{b} X \to e D^{*\pm}\mu X) = 160 \pm 37 \rm{(stat.)}^{+ 30}_{- 57}\rm{(syst.)}~\rm{pb}.
\end{equation}
This includes both unlike- and like-sign $D^*\mu$ combinations.
The leading-order cross sections predicted by {\sc Pythia} and {\sc
  Herwig} in the same kinematic range are $\sigma_{\rm vis}(ep \to e b
\bar{b} X \to e D^*\mu X) =$ 80 and 38 pb, respectively.  The measured
cross section is larger than, but compatible with, the
FMNR$\otimes${\sc Pythia} NLO prediction
\begin{equation}
\sigma_{\rm vis}^{\rm NLO}(ep \to e b \bar{b} X \to e D^{*\pm}\mu X) = 67^{+20}_{-11} \rm{(NLO)}^{+ 13}_{- 9}\rm{(frag. \oplus br.)}~\rm{pb},       
\end{equation}
where the first error refers to the uncertainties of the FMNR
parton-level calculation, and the second error refers to the
uncertainties related to fragmentation and decay.

For the photoproduction subsample, a visible cross section in the
kinematic range $Q^2 < 1$ GeV$^2$ and $0.05 < y < 0.85$ was obtained:
\begin{equation}
\label{sig:PHP}
\sigma_{{\rm vis},\gamma p}(ep \to e b \bar{b} X \to e D^{*\pm}\mu X) = 115 \pm 29 \rm{(stat.)}^{+21}_{-27}\rm{(syst.)}~\rm{pb}.
\end{equation}

This can be compared to the NLO prediction from FMNR$\otimes${\sc Pythia},
\begin{equation}
\sigma_{{\rm vis},\gamma p}^{\rm NLO}(ep \to e b \bar{b} X \to e D^{*\pm}\mu X) = 54^{+15}_{-10} \rm{(NLO)}
^{+10}_{-7}\rm{(frag. \oplus br.)}~\rm{pb}.
\end{equation}

As in the inclusive case, the NLO prediction underestimates the
measured cross section by about a factor of 2, but is compatible with
the measurement (Table \ref{tab:cross}).

From the DIS sample, a visible cross section in the kinematic range
$Q^2 > 2$ GeV$^2$, $0.05 < y < 0.7$ and $p_T^{D^*} > 1.5$ GeV (other
$D^*$ and muon cuts as for Eq.~(\ref{sig:tot})) of
\begin{equation}
\label{sig:DIS}
\sigma_{\rm vis,DIS}(ep \to e b \bar{b} X \to e D^*\mu X) = 58 \pm 29 \rm{(stat.)}^{+11}_{-20}\rm{(syst.)} ~\rm{pb}
\end{equation}
was obtained.

Again, the cross sections obtained from {\sc Rapgap} (used to compute
acceptance corrections for the central signal extraction) and {\sc Herwig} 
(used for systematic checks, particularly with regard to
differences in the $b\bar{b}$ correlations) in the same kinematic
regime are considerably lower, $\sigma(ep \to e b \bar{b} X \to e
D^*\mu X) =$ 26 and 10 pb, respectively.  An NLO prediction is not
available for this kinematic region.

\subsection{Comparison to H1 results}

A photoproduction cross section similar to Eq.~(\ref{sig:PHP}) in a
slightly different kinematic range, $p_T^{D^*} > 1.5$ GeV,
$|\eta^{D^*}| < 1.5$, $p^{\mu} > 2.0$ GeV, $|\eta^{\mu}| < 1.735$,
$Q^2 < 1$ GeV$^2$ and $0.05 < y < 0.75$ has been obtained by the H1
collaboration \cite{pl:b621:56-71}
\begin{equation}
\sigma_{{\rm vis},\gamma p}^{\rm H1}(ep \to e b \bar{b} X \to e D^{*\pm}\mu X;\rm{H1}) = 206 \pm 53 \rm{(stat.)} \pm {35} \rm{(syst.)}~\rm{pb}.       
\end{equation}
The ZEUS cross section of Eq.~(\ref{sig:PHP}) extrapolated to the 
same kinematic range as the H1 measurement
using FMNR$\otimes${\sc Pythia} is
\begin{equation}
\sigma_{{\rm vis},\gamma p}(ep \to e b \bar{b} X \to e D^{*\pm}\mu X;\rm{H1}) = 135 \pm 33 \rm{(stat.)}^{+ 24}_{- 31}\rm{(syst.)}~\rm{pb},
\end{equation}
which is somewhat smaller, but in agreement within errors.

The corresponding FMNR$\otimes${\sc Pythia} NLO prediction is
\begin{equation}
\sigma_{{\rm vis},\gamma p}^{\rm NLO}(ep \to e b \bar{b} X \to e D^{*\pm}\mu X ; \rm{H1}) = 61^{+17}_{-12} \rm{(NLO)}
^{+12}_{-8}\rm{(frag. \oplus br.)}~\rm{pb}.
\end{equation}
This is larger than the NLO cross section evaluated by H1
\cite{pl:b621:56-71} due to the inclusion of the hadron-like photon
contribution, the inclusion of secondary-muon branching fractions for
$D^*$ and $\mu$ from the same $b$ quark (Table \ref{tab:branchings}),
and a detailed simulation of the kinematics of the $b \to B \to D^*$
chain rather than direct collinear fragmentation of $b$ quarks into
$D^*$ mesons. The data to NLO ratio is consistent with the results in
the ZEUS kinematic range.

\subsection{Cross sections for $D^*\mu$ from the same $b$ quark}

In all the cross sections evaluated above, a significant part of the
systematic error arises from the fraction of the beauty contribution
in the charm enriched or like-sign regions, where it cannot be well
measured ($\Delta R > 2$ region in Fig.~\ref{fig2}a; Figs.~\ref{fig2}b
and \ref{fig2}d). This fraction depends on details of the description
of $b\bar b$ correlations in the MC used for the signal extraction.
In the beauty-enriched low-$\Delta R$ unlike-sign region, which
dominates the fit of the beauty fraction, about 95\% of the $D^*\mu$
pairs are produced from the same parent $b$ quark.  The systematic
error can thus be reduced by reinterpreting the measurements in terms
of cross sections for this subprocess only.  The corresponding cross
section for photoproduction of a $D^*$ and $\mu$ from the same $b$
quark (always unlike sign, same kinematic cuts as for
Eq.~(\ref{sig:PHP})) is
\begin{equation}
\label{sig:PHPsame}
\sigma_{{\rm vis},\gamma p}(ep \to e b (\bar{b}) X, b (\bar{b}) \to e D^*\mu X) =  52 \pm 13 \rm{(stat.)}^{+9}_{-11}\rm{(syst.)} ~\rm{pb}
\end{equation}
where $b (\bar{b})$ stands for the sum of $b$ and $\bar b$ cross
sections, and all other cuts remain the same.

This can be compared with the NLO prediction
\begin{equation}
\sigma_{{\rm vis},\gamma p}^{\rm NLO}(ep \to e b (\bar{b}) X, b (\bar{b}) \to e D^*\mu X) 
=  29^{+8}_{-5} \rm{(NLO)}^{+5}_{-4}\rm{(frag. \oplus br.)}~\rm{pb}.
\end{equation}

For the DIS kinematic range (same as Eq.~(\ref{sig:DIS}))
\begin{equation}
\label{sig:DISsame}
\sigma_{\rm vis,DIS}(ep \to e b (\bar{b}) X,  b (\bar{b}) \to e D^*\mu X) = 28 \pm 14 \rm{(stat.)}^{+5}_{-10}\rm{(syst.)} ~\rm{pb}
\end{equation}
is obtained.

\subsection{Parton-level cross sections}

For a direct comparison with the NLO parton-level predictions, the
measured visible cross sections were extrapolated to $b$-quark level.
In order to minimize the systematic error, the $b$-level cross section
is quoted for individual $b$ (or $\bar{b}$) production rather than for
correlated $b\bar{b}$-pair production, i.e. using the cross sections
displayed in Eqs.~(\ref{sig:PHPsame}) and (\ref{sig:DISsame}).

A significant fraction of the parent $b$ quarks of the selected events
is expected to have very low $p_T^b$ values
\cite{thesis:longhin:2004}. Therefore, cross sections with no cut on
$p_T^b$ have been measured.  Furthermore, there is a strong
correlation between the pseudorapidity of the $D^*\mu$ system and the
rapidity of the parent $b$ quark, $\zeta^b =
\frac{1}{2}\ln\frac{E_b+p_{z,b}}{E_b-p_{z,b}}$.  In order to reflect
the limited angular acceptance of the detector for both the $D^*$ and
the muon, the cross-section measurement was restricted to $\zeta^b <
1$.  In this range, restricted to photoproduction, the $p_T^b$ and
$\zeta^b$ distributions of {\sc Pythia} (after parton showering) agree
with the central NLO $b$-quark spectra from FMNR to within $\pm15$\%
\cite{thesis:longhin:2004}.  Therefore, {\sc Pythia} was used to
extrapolate the visible cross section for the photoproduction region.
Similarly, the corresponding {\sc Rapgap} spectra for the DIS case
agree \cite{thesis:longhin:2004} with the central NLO predictions from
HVQDIS.

The acceptance for $b$ quarks due to the kinematic cuts on the
fragmentation and decay products ranges from $\sim$ 4\% at $p_T^b=0$
GeV to $\sim$ 55\% at $p_T^b=10$ GeV. The remaining part of the
extrapolation is due to the relevant branching ratios.

The extrapolation implies additional systematic uncertainties from the
$b$-quark fragmentation (+5/-15\%) and decay ($\pm 9$\%) and the
details of the shape of the $p_T^b$ spectrum ($\pm 5$\%).  The
extrapolation was calculated assuming the validity of the NLO $p_T^b$
shape and is therefore valid only in the context of this theoretical
framework; the uncertainty for the visible cross section corresponding
to a potential deviation from this shape, namely the reweighting of
the $p_T^{D^*\mu}$ spectrum, is removed.  The result for the
extrapolated cross section for $\zeta^b<1,\ Q^2<1~{\rm GeV}^2,\
0.05<y<0.85$ and $m_b = 4.75$ GeV was
\begin{equation}
\label{sig:PHPext}
\sigma_{\gamma p}(ep\to b (\bar{b})X)= 11.9 \pm 2.9\rm{(stat.)}^{+1.8}_{-3.3}\rm{(syst.)}~\rm{nb}.
\end{equation}
The corresponding result for the extrapolated cross section for
$\zeta^b<1,\ Q^2>2~{\rm GeV}^2,\ 0.05<y<0.7$ and $m_b = 4.75$ GeV was
\begin{equation}
\label{sig:DISext}
\sigma_{DIS}(ep\to b (\bar{b})X)= 3.6 \pm 1.8\rm{(stat.)}^{+0.5}_{-1.4}\rm{(syst.)}~\rm{nb}.
\end{equation}
These cross sections are to be compared to the NLO prediction for the same kinematic
range using the FMNR calculation of
\begin{equation}
\sigma_{\gamma p}^{NLO}(ep\to b (\bar{b)}X) =  5.8^{+ 2.1}_{-1.3}{\rm ~nb},
\end{equation}

and to the NLO HVQDIS prediction of
\begin{equation}
\sigma_{DIS}^{NLO}(ep\to b (\bar{b)}X) =  0.87^{+ 0.28}_{-0.16}{\rm ~nb}.
\end{equation}

These cross sections are presented in Fig.~\ref{fig7}.  The ratio of
measured to predicted cross sections in the photoproduction region
remains the same as the ones obtained from the comparison at visible
level (Table \ref{tab:cross}).  This confirms the self-consistency of
the extrapolation procedure used.

\subsection{Comparison to previous ZEUS measurements}
   
In order to compare the photoproduction cross section to previous ZEUS
results \cite{pl:b467:156,*epj:c18:625,pr:d70:012008}, the cross
sections Eq.~(\ref{sig:PHPsame}) or equivalently
Eq.~(\ref{sig:PHPext}), already referring to the production of a
single $b$ quark, need to be translated into a differential cross
section, $\frac{d\sigma}{dp_{T}^b}$, in the pseudorapidity range
$|\eta^b| < 2$\cite{pr:d70:012008}.  The median $p_{T}^b$ value for
events satisfying the cuts for Eq.~(\ref{sig:PHPsame}) is 6.5 GeV
\cite{thesis:longhin:2004}.  The measured cross section,
Eq.~(\ref{sig:PHPsame}), is therefore extrapolated to this value using
FMNR$\otimes${\sc Pythia}, yielding
\begin{equation}
\label{sig:PHPdiff}
\frac{d\sigma}{dp_{T}^{b}}(p_{T}^b = 6.5 {\rm ~GeV}, |\eta^b| < 2)
= 0.30 \pm 0.07\rm{(stat.)}^{+0.05}_{-0.06}\rm{(syst.)}~\rm{nb}.
\end{equation}

This result is compared to theory and previous measurements in
Fig.~\ref{fig9}.  It is higher than, but consistent with, these
measurements.

\section{Conclusions}

Cross sections for beauty production in $ep$ collisions at HERA have
been measured in both the photoproduction and DIS regimes using an
analysis technique based on the detection of a muon and $D^*$.
Agreement is obtained with the corresponding H1 result.  Since the
analysis is sensitive to $b$-quark production near the kinematic
threshold, the measured visible cross sections were extrapolated to
$b$-quark cross sections without an explicit cut on $p_T^b$. Both at
visible and at quark level, the measured cross sections exceed the NLO
QCD predictions, but are compatible within the errors.  The data to
NLO ratio is also larger than, but compatible with, previous ZEUS
measurements of the $b$ production cross section at higher $p_T^b$.

\section*{Acknowledgements}
We thank the DESY Directorate for their strong support and
encouragement.  The remarkable achievements of the HERA machine group
were essential for the successful completion of this work and are
greatly appreciated.  We are grateful for the support of the DESY
computing and network services.  The design, construction and
installation of the ZEUS detector have been made possible owing to the
ingenuity and effort of many people who are not listed as authors.  It
is also a pleasure to thank S. Frixione for help with the theoretical
predictions.

\providecommand{\etal}{et al.\xspace}
\providecommand{\coll}{Coll.\xspace}
\catcode`\@=11
\def\@bibitem#1{%
\ifmc@bstsupport
  \mc@iftail{#1}%
    {;\newline\ignorespaces}%
    {\ifmc@first\else.\fi\orig@bibitem{#1}}
  \mc@firstfalse
\else
  \mc@iftail{#1}%
    {\ignorespaces}%
    {\orig@bibitem{#1}}%
\fi}%
\catcode`\@=12

\begin{table}[hbt!]
\begin{center}
\begin{tabular}{|c|c|c|c|c|c|c|}
\hline
\multirow{2}{*}{Sample} & \multirow{2}{*}{Cuts} & \multirow{2}{*}{Data} & \multirow{2}{*}{Beauty} & \multicolumn{2}{c|}{Charm}  & \multirow{2}{*}{Fake $D^*$}\\
       &      &      &        & prompt $\mu$ & fake $\mu$   &            \\
\hline
\multicolumn{2}{|c|}{~} &\multicolumn{5}{c|}{Inclusive sample} \\
\hline
\multirow{2}{*}{\mbox{unlike sign}} & $\Delta R < 2$, $3<M<5$ GeV & 41 & 28.8 & 3.6 & 2.5 & 11.2\\
~                                         & $\Delta R > 2$              & 93 & 6.5 & 56.5 & 18.4 & 6.4\\
\hline
\multirow{2}{*}{like sign}   & $\Delta R < 2$, $3<M<5$ GeV & 18  & 1.6 &  0.7 & 1.3 & 9.2 \\
~ &                                         $\Delta R > 2$              & 36 & 11.1 &  2.0 & 21.2 & 5.2\\
\hline

\multicolumn{2}{|c|}{~} &\multicolumn{5}{c|}{$\gamma p$} \\
\hline
\multirow{2}{*}{\mbox{unlike sign}} & $\Delta R < 2$, $3<M<5$ GeV & 31 &  23.6 &  1.2 &  1.6 & 8.3\\
~                                         & $\Delta R > 2$              & 79 &  6.2 &  48.8 &  14.0 & 6.2\\
\hline
\multirow{2}{*}{like sign}   & $\Delta R < 2$, $3<M<5$ GeV &  14 & 1.5 &  0.6 &  0.4 & 6.9\\
~ &                                         $\Delta R > 2$              &  28 & 9.1 &  1.9 &  17.1 & 5.2\\
\hline

\multicolumn{2}{|c|}{~} &\multicolumn{5}{c|}{DIS} \\
\hline
\multirow{2}{*}{\mbox{unlike sign}} & $\Delta R < 2$, $3<M<5$ GeV & 11 &  8.1 &  1.6 &  0.5 & 1.0\\
~                                         & $\Delta R > 2$              & 14 &  3.2 &  6.1 &  1.2 & 3.9\\
\hline
\multirow{2}{*}{like sign}   & $\Delta R < 2$, $3<M<5$ GeV &  3 & 1.2  &  0.1 &  0.5 & 0.3\\
~ &                                         $\Delta R > 2$              &  10 & 5.0 &  0.2 &  2.6 & 1.2\\
\hline

\end{tabular}
\end{center}
\caption{Composition of the final $D^*\mu$ event samples (number of events)
as determined from the fit to the $\Delta R$ distribution.}
\label{tab_evts}
\end{table}

\begin{table}[hbt!]
\begin{center}
\begin{tabular}{|c|c|}
\hline
channel & branching fraction w/o $B^0$-$\bar B^0$ mixing \\
\hline
\multirow{2}{*}{$b \to D^{* \pm}$ inclusive} & $17.3 \pm 2.0$ \% \\
 & $86\pm3$ \% $D^{*+}$, $14\pm3$ \% $D^{*-}$ \\ 
\hline
 $b \to \mu^-$ direct & $10.95 \pm 0.27$ \% \\
 $b \to \mu^{+}$ indirect & $8.27 \pm 0.40$ \% \\
 $b \to \mu^{-}$ indirect & $2.21 \pm 0.50$ \% \\
\hline
all $b \to \mu^{\pm} $ & $21.43 \pm 0.70$ \% \\
\hline
$ b \bar b \to D^{*\pm} \mu^{\pm} $ (diff.~$b$'s) & $4.34 \pm 0.92$ \% \\
$ b \bar b \to D^{*\pm} \mu^{\mp} $ (diff.~$b$'s) & $3.08 \pm 0.60$ \% \\
\hline
$b \to D^{*+} \mu^- $ direct & $2.75 \pm 0.19$ \%  \\
$b \to D^{*\pm} \mu^{\mp} $ indirect  & $1.09 \pm 0.27$ \% \\
\hline
all $b \to D^{*\pm} \mu^{\mp} $  & $3.84 \pm 0.33$ \% \\
\hline
\end{tabular}
\end{center}
\caption{Branching fractions assumed for cross-section determinations.
  The indirect contributions include cascade decays into muons via
  charm, anticharm, $\tau^\pm$ and $J/\psi$.  The values in the Table
  are given before the inclusion of the effect of $B^0$-$\bar B^0$
  mixing (mixing parameter $\chi = 0.1257 \pm 0.0042$) \protect\cite{pl:b592:1}.}
\label{tab:branchings}
\end{table}

\begin{table}[hbt!]
\begin{center}
\begin{tabular}{|c|c|c|c|}
\hline
cross section & measured & NLO QCD & ratio \\
\hline
inclusive, visible, Eq.(\ref{sig:tot}) 
  & $160 \pm 37 ^{+ 30}_{- 57}~\rm{pb}$
  & $67^{+24}_{-14}~\rm{pb}$
  & $2.4^{+0.9}_{-1.3}$ \\
$\gamma p$, visible, Eq.(\ref{sig:PHP}) 
  & $115 \pm 29 ^{+21}_{-27}~\rm{pb}$
  & $54^{+18}_{-12}~\rm{pb}$
  & $2.1^{+0.8}_{-1.0}$ \\
DIS, visible, Eq.(\ref{sig:DIS})
  & $ 58 \pm 29 ^{+11}_{-20}~\rm{pb}$
  & -
  & - \\
\hline
$\gamma p $, vis. same $b$, Eq.(\ref{sig:PHPsame})  
  & $ 52 \pm 13 ^{+9}_{-11}~\rm{pb}$
  & $ 29^{+9}_{-6}~\rm{pb}$
  & $1.8^{+0.7}_{-0.8}$ \\
DIS, vis. same $b$, Eq.(\ref{sig:DISsame})
  & $28 \pm 14 ^{+5}_{-10}~\rm{pb}$
  & - 
  & -  \\
\hline
$\gamma p$, $b$ quark, Eq.(\ref{sig:PHPext})
  & $11.9 \pm 2.9 ^{+1.8}_{-3.3}~\rm{nb}$
  & $5.8^{+2.1}_{-1.3}~{\rm nb}$
  & $2.0^{+0.8}_{-1.1}$ \\
DIS, $b$ quark, Eq.(\ref{sig:DISext})
  & $3.6 \pm 1.8 ^{+0.5}_{-1.4}~\rm{nb}$
  & $0.87^{+0.28}_{-0.16}~{\rm nb}$
  & $4.2^{+2.3}_{-2.9}$ \\
\hline
$\gamma p$, $b$ differential, Eq.(\ref{sig:PHPdiff})
  & $0.30 \pm 0.07 ^{+0.05}_{-0.06}~\rm{nb/GeV} $
  & $0.16^{+0.04}_{-0.02}~{\rm nb/GeV}$
  & $1.8^{+0.7}_{-0.8}$ \\
\hline
\end{tabular}
\end{center}
\caption{Comparison of measured and predicted cross sections.  For the
  measured cross sections, the first error is statistical, and the
  second systematic.  For the QCD prediction, the error is due to the
  parton-level NLO calculation convoluted with the uncertainties of
  fragmentation and decay to the visible final state.  The number in
  parentheses refers to the corresponding equation for each cross
  section, see text.  For the definition of the kinematic range of
  each cross section see text.}
\label{tab:cross}
\end{table}


\setlength{\unitlength}{1mm}
\begin{figure}[ht]
  \begin{center}
    \begin{picture}(150,140)
      \put(0,70){\epsfig{file=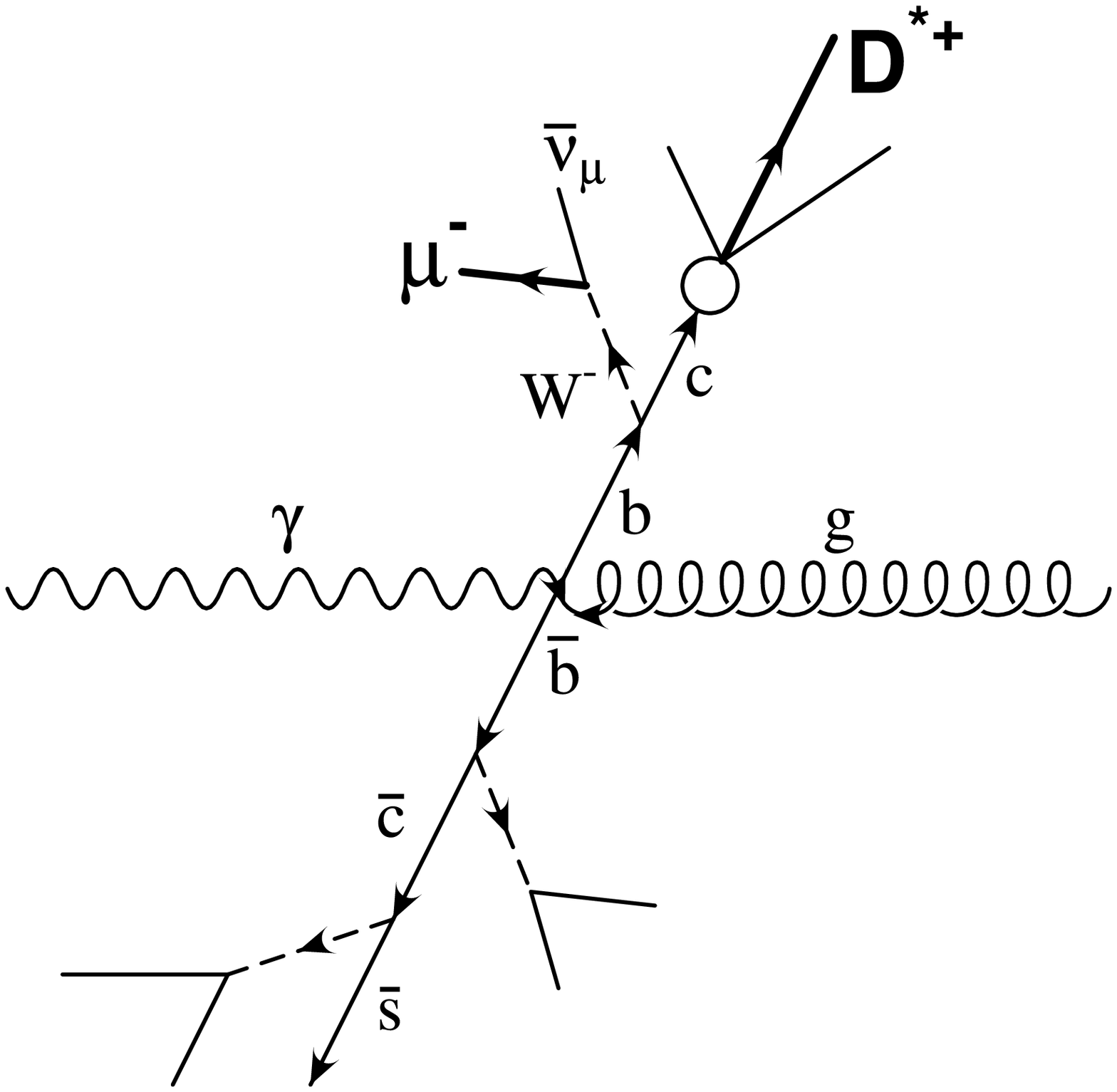,width=6cm,clip}}
      \put(80,70){\epsfig{file=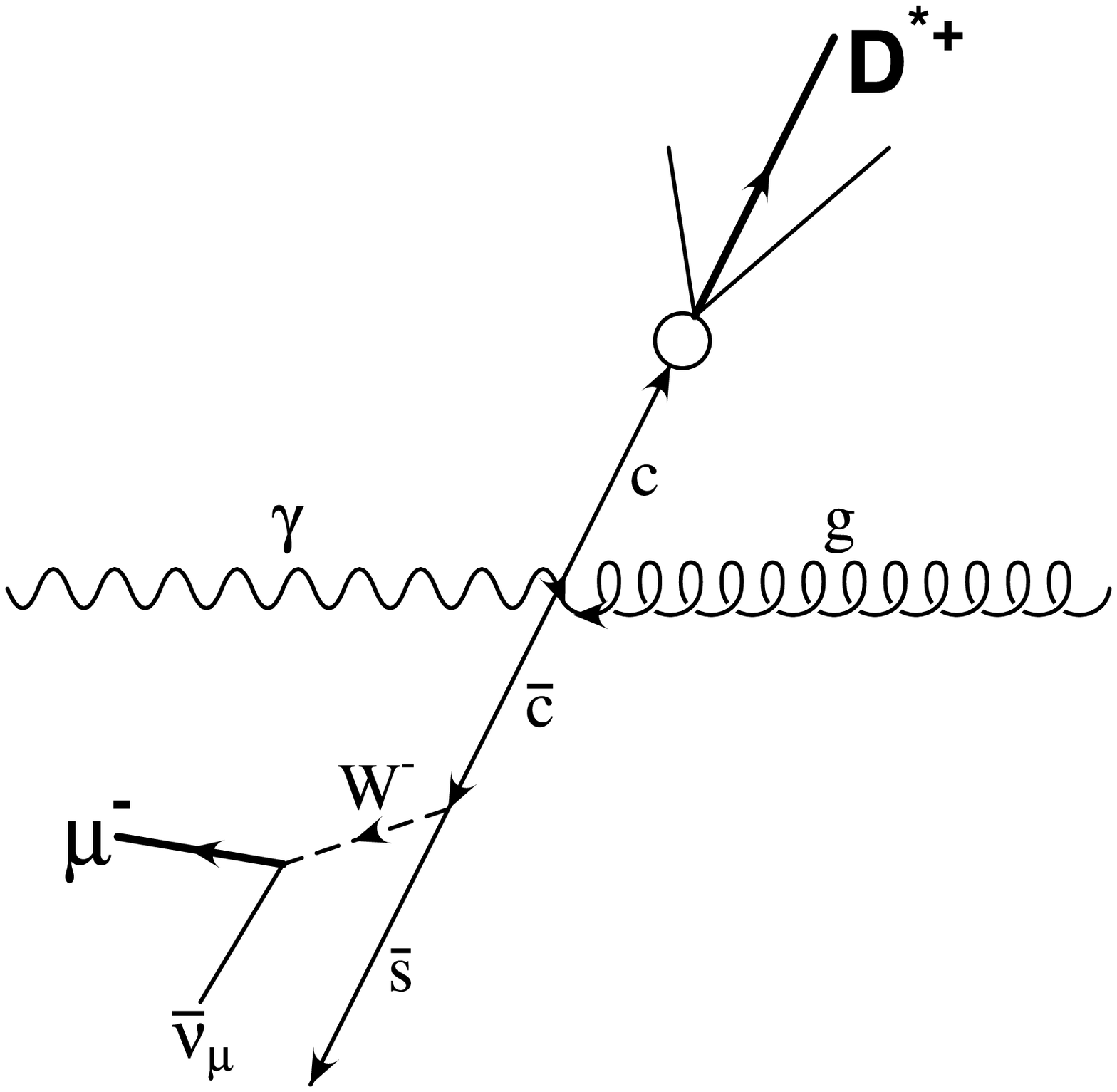,width=6cm,clip}}
      \put(0,0){\epsfig{file=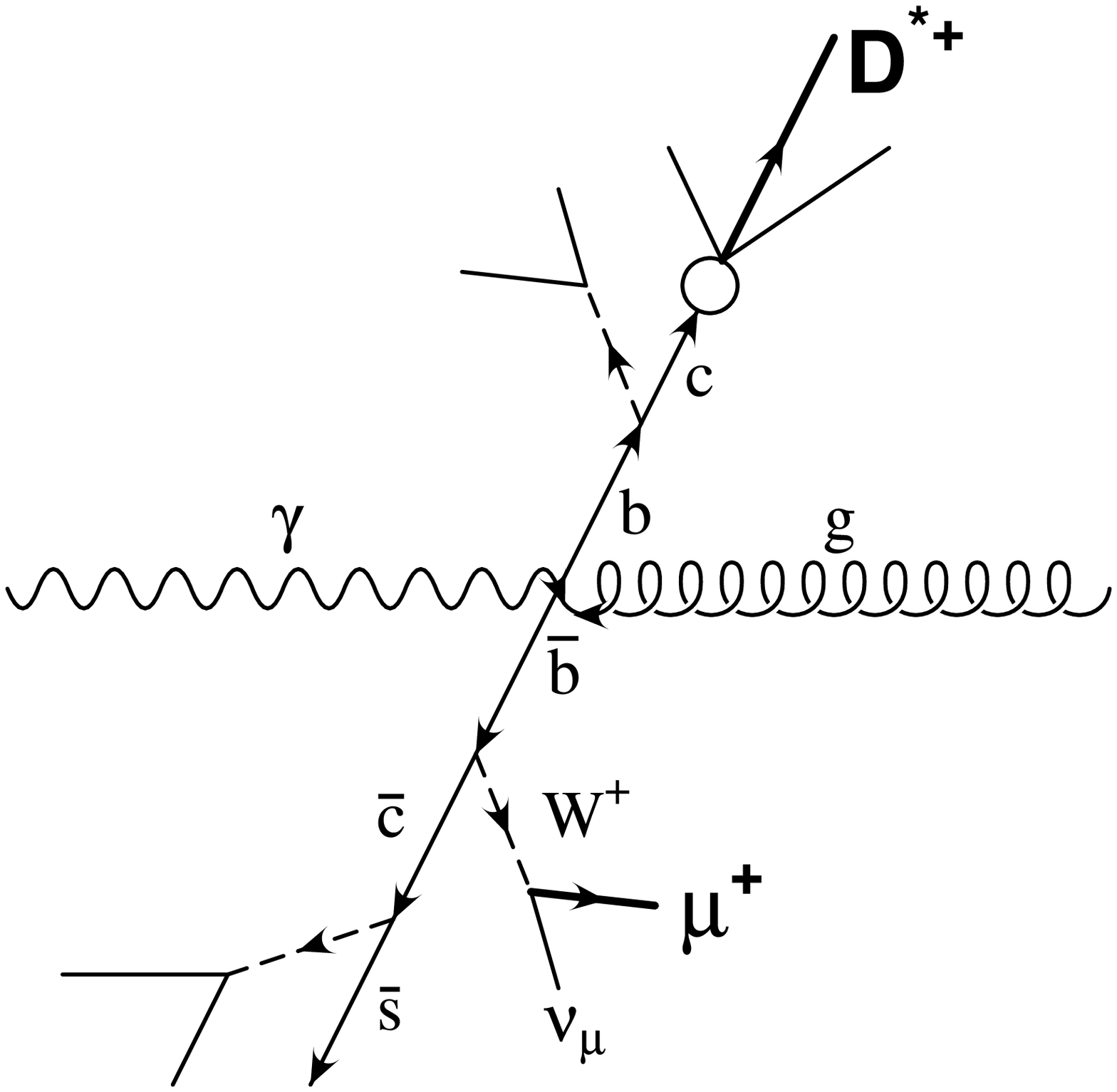,width=6cm,clip}}
      \put(80,0){\epsfig{file=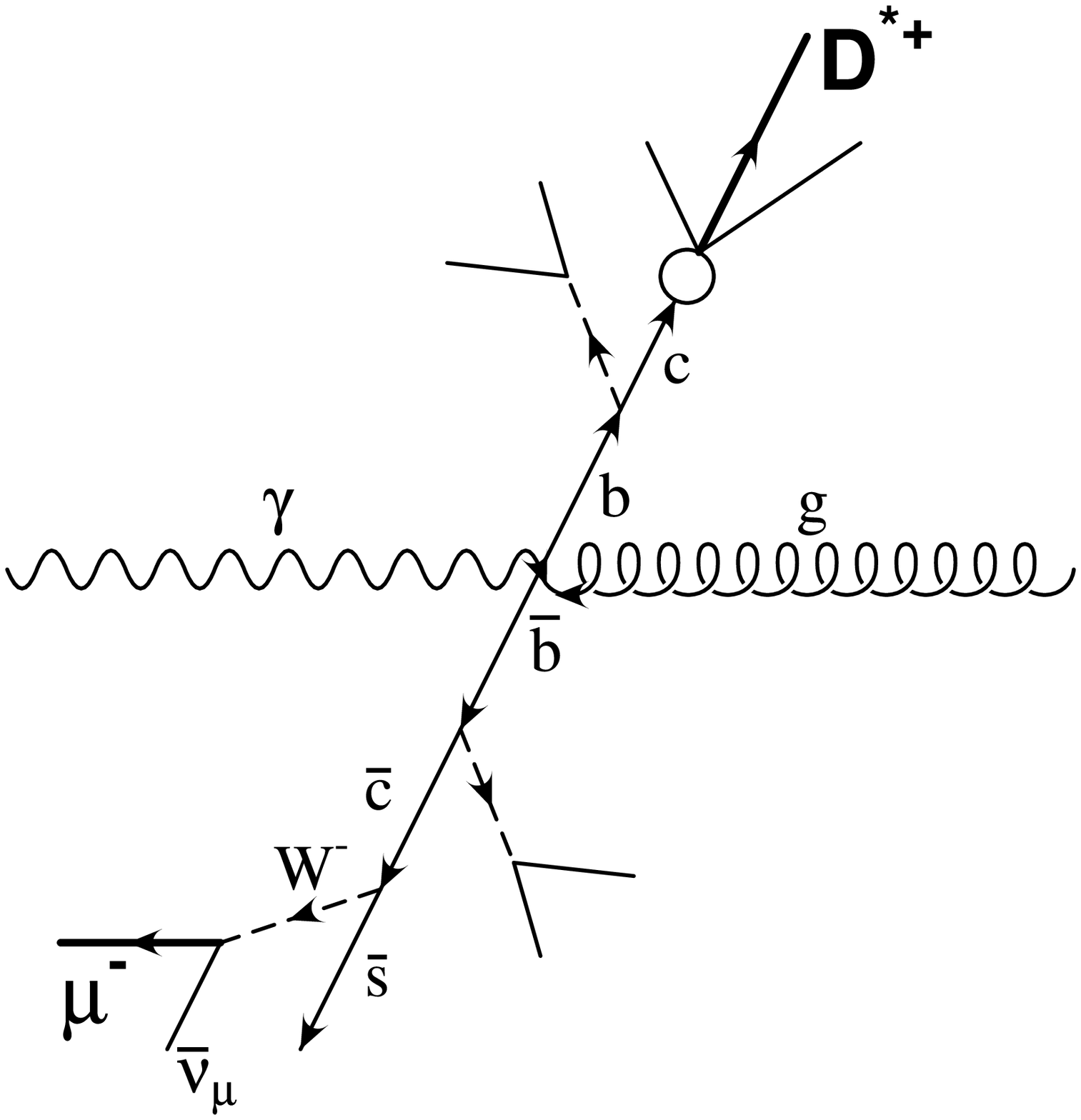,width=6cm,clip}}
      \put(5,120){(a)}
      \put(85,120){(b)}
      \put(5,50){(c)}
      \put(85,50){(d)}
    \end{picture}
\vspace{1cm}
  \end{center}
  \caption{Processes leading to $D^*\mu$ final states.}
  \label{fig_topol}
\end{figure}

\begin{figure}[ht]
\begin{center}
\includegraphics[totalheight=12cm]{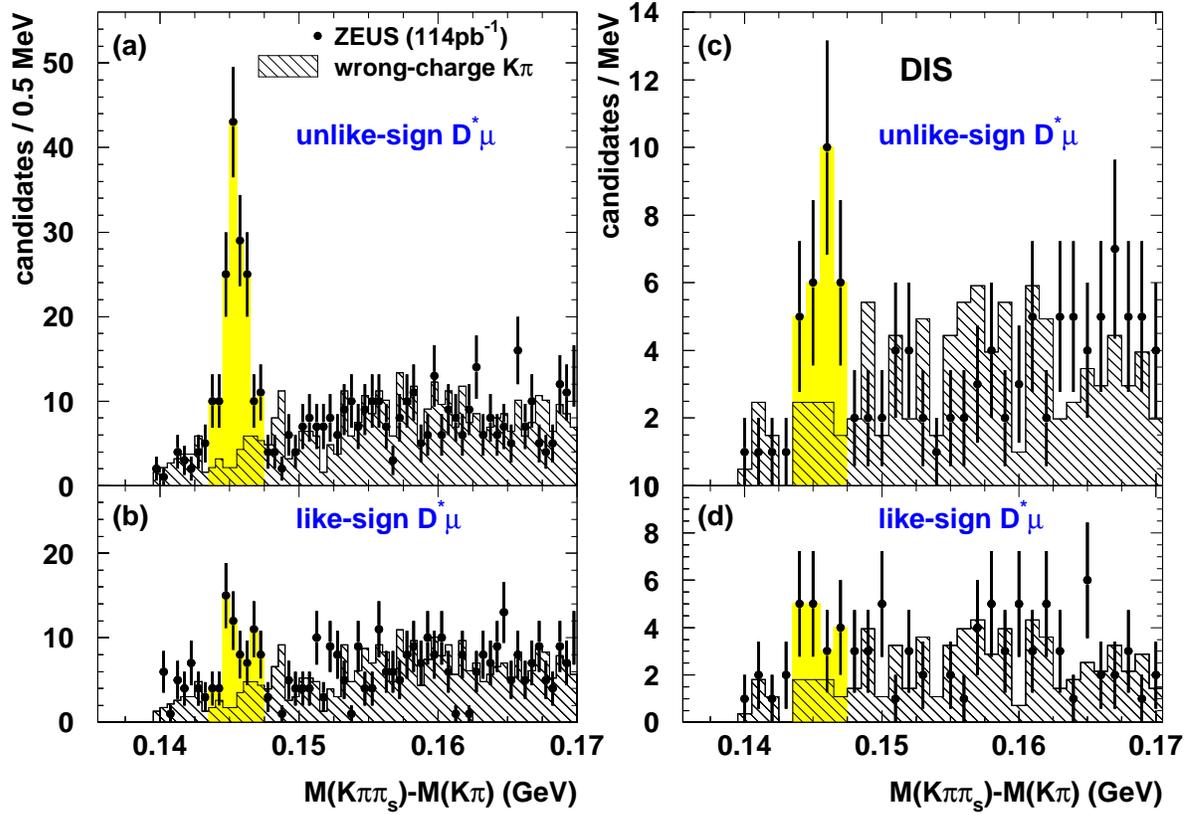}
\end{center}
\caption{Distribution of $\Delta M$ for data (full circles) and
  combinatorial background (hatched histogram) for (a) inclusive
  unlike-sign (b) inclusive like-sign (c) DIS unlike-sign and (d) DIS
  like-sign muon-$D^*$ combinations. The $D^*$ signal region is
  indicated by the shaded area.}
\label{fig1}
\end{figure}

\begin{figure}[ht]
\begin{center}
\includegraphics[totalheight=16cm]{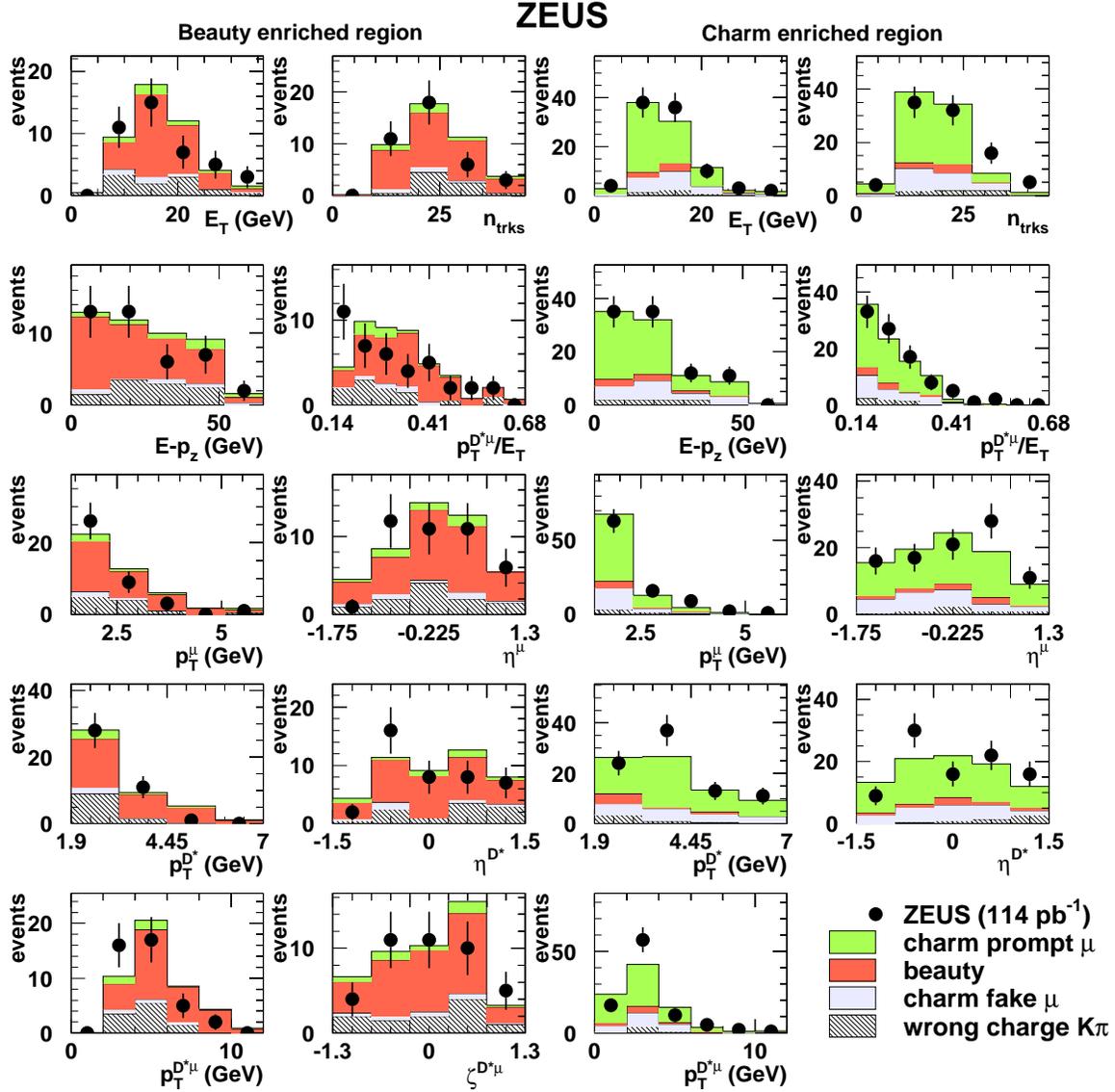}
\end{center}
\caption{Distributions of variables used in the event selection or
  relevant for the event kinematics, for the unlike-sign inclusive
  sample after all cuts.  All variables are defined in the text except
  $n_{trks}$ (total number of tracks) and $\zeta^{D^*\mu}$ (rapidity
  of the $D^*\mu$ system).  The distributions are shown separately for
  the beauty- and charm-enriched regions as defined in Section
  \ref{sect:signal}.  The beauty, charm, fake-muon and wrong-charge
  $K\pi$ background distributions are indicated by different shading
  styles, and normalized to the fractions determined later in the
  analysis.}
\label{fig:controlplots}
\end{figure}

\begin{figure}[ht]
\begin{center}
\includegraphics[totalheight=10.5cm]{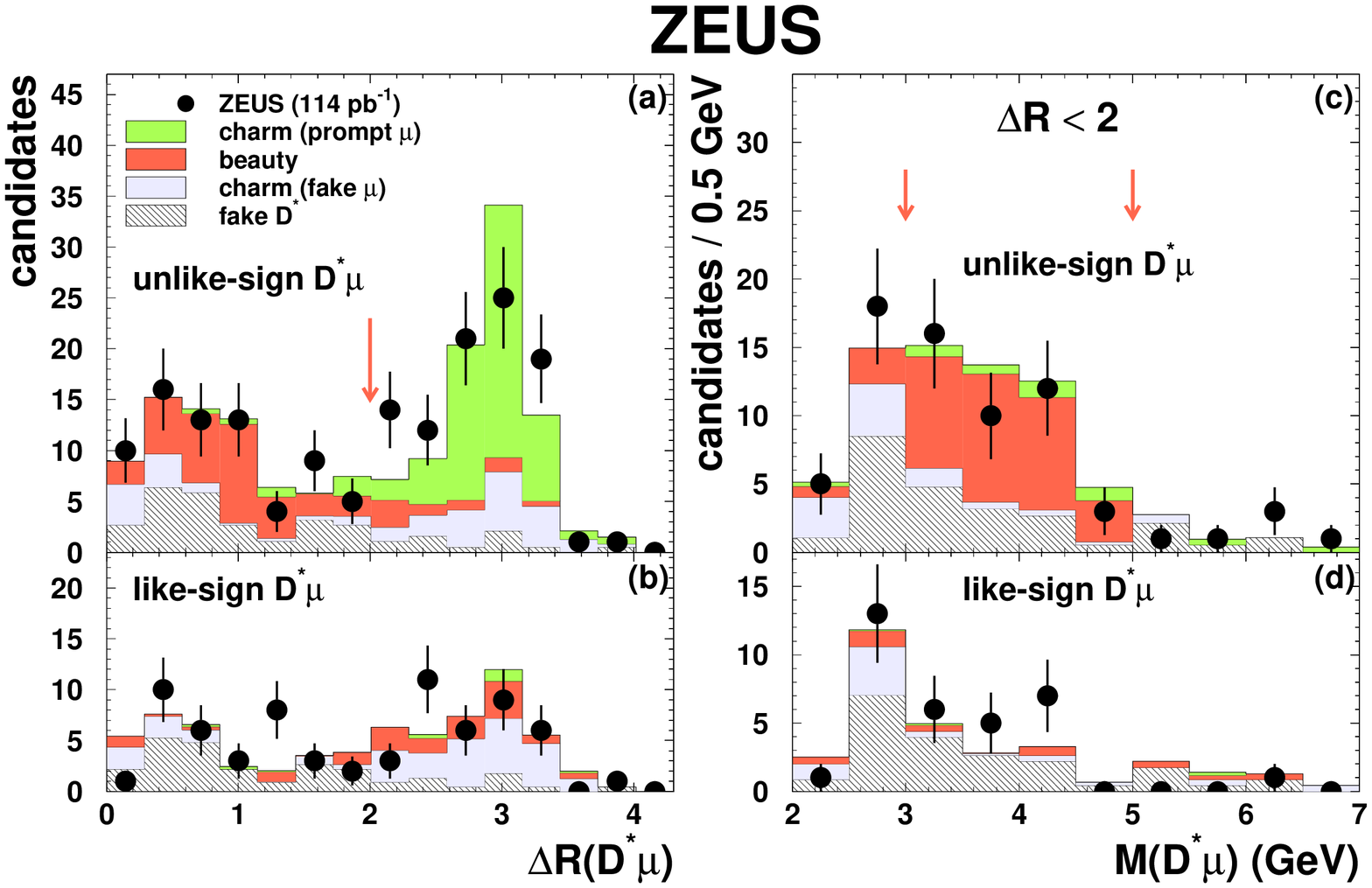}
\end{center}
\caption{Distribution of (a),(b) $\Delta R(D^*\mu)$ and (c),(d)
  $M(D^*\mu)$ for inclusive data (full circles), beauty and charm
  signal, fake-$\mu$ and fake-$D^*$ backgrounds. The latter are
  distinguished by different shading styles.  Unlike-sign and
  like-sign $D^*\mu$ combinations are shown separately. Cuts described
  in the text are indicated by the arrows. The relative contributions of charm
  and beauty are determined by the fit to the $\Delta R$ distribution.
}
\label{fig2}
\end{figure}

\begin{figure}[ht] \begin{center}
\includegraphics[totalheight=16cm]{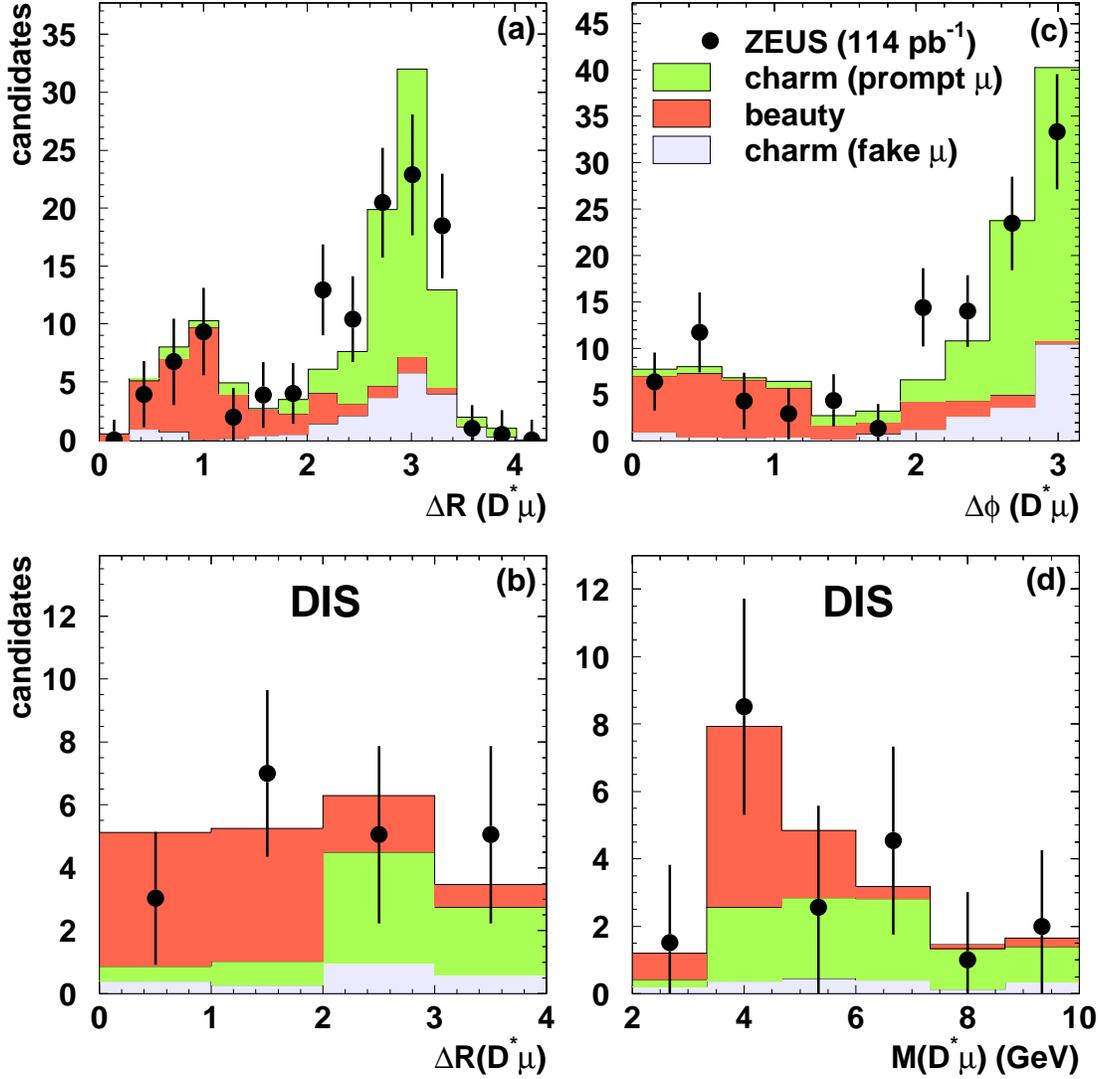} 
\end{center} 
\caption{Distributions of (a) $\Delta R(D^*\mu)$, (c) $\Delta
  \phi(D^*\mu)$ for unlike-sign events in the inclusive sample, (b)
  $\Delta R(D^*\mu)$ and (d) $M(D^*\mu)$ for unlike-sign events in the
  DIS sample after subtraction of the fake-$D^*$ background. Data
  points (full circles) are shown together with the contributions from
  beauty and charm, as determined in the fit. The fake-muon
  contribution from charm is shown separately, but fitted together
  with the charm contribution.}
\label{fig3}
\end{figure}

\begin{figure}[ht]
\begin{center}
\includegraphics[totalheight=14cm]{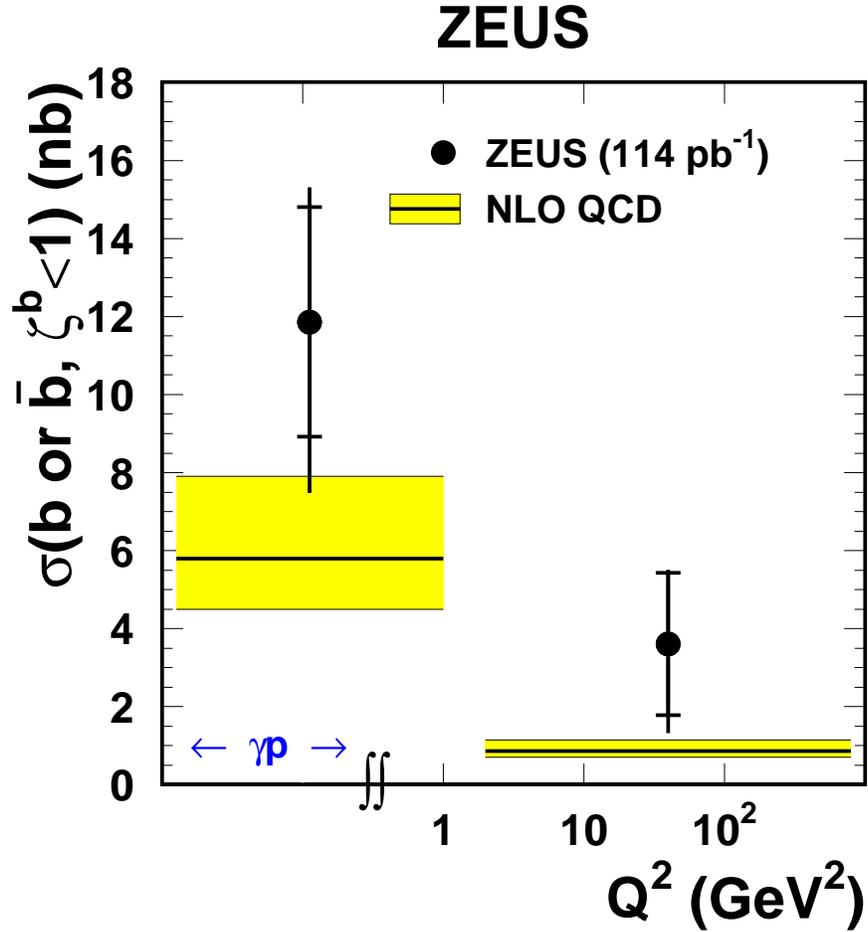}
\end{center}
\caption{Cross section for single $b$ or $\bar b$-quark production in
  the rapidity range $\zeta^b < 1$ for photoproduction (left) and DIS
  (right), compared to NLO QCD predictions from FMNR (left) and HVQDIS
  (right).  The $\gamma p$ cross section is for $0.05 < y < 0.85$,
  $Q^2<1 \gev^2$ and the DIS cross section for $0.05 < y < 0.7$, $Q^2
  > 2 \gev^2$.  The cross sections are integrated over the $Q^2$
  ranges, and over the full $p_T^b$ range.}
\label{fig7}
\end{figure}

\begin{figure}[ht]
\begin{center}
\includegraphics[totalheight=11cm]{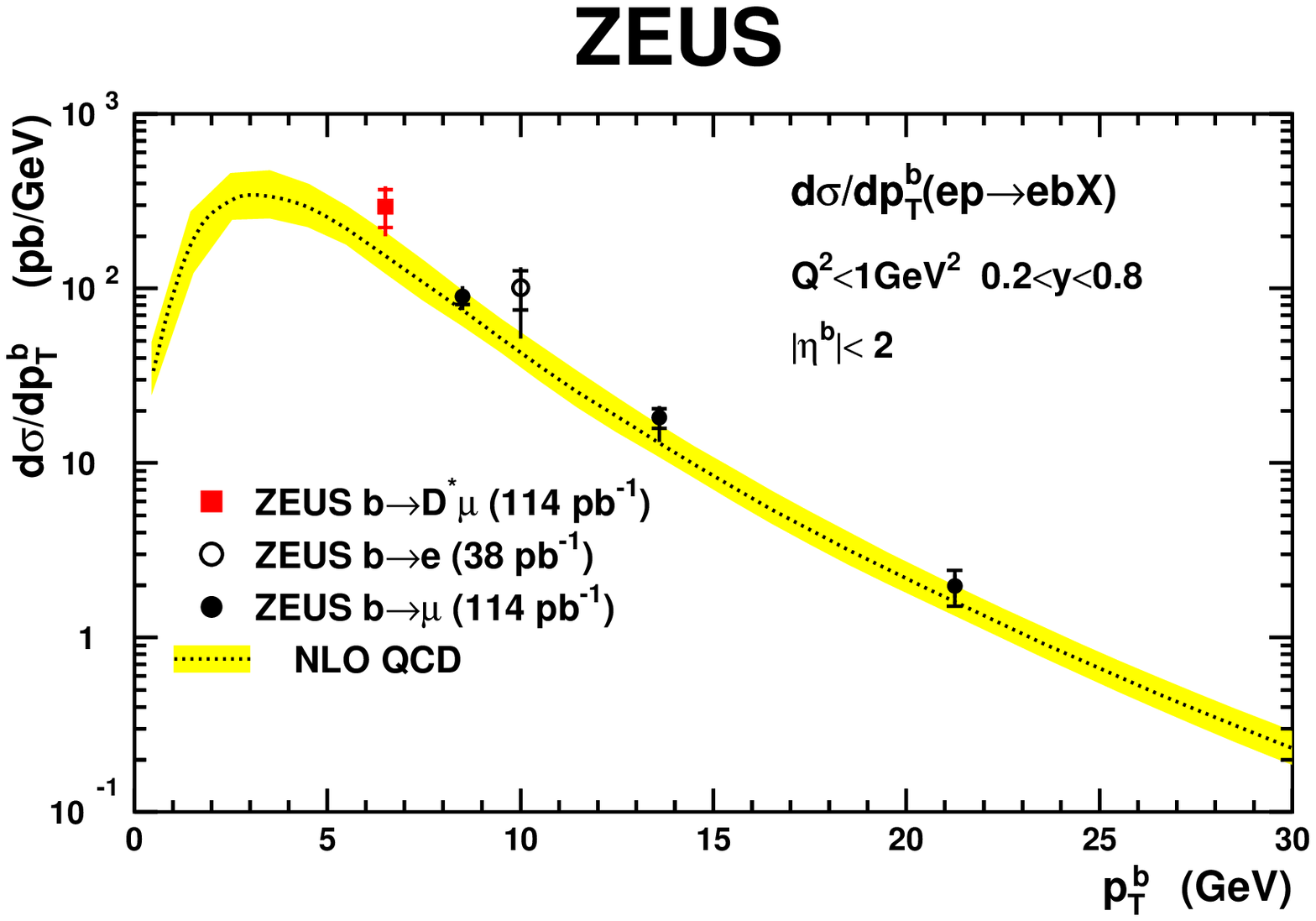}
\end{center}
\caption{Differential cross-section $\frac{d\sigma}{dp_T^b}$ of this
  analysis (filled square) compared to measurements from previous ZEUS
  analyses, after rescaling to the kinematic range indicated in the
  Figure.}
\label{fig9}
\end{figure}
\end{document}

%% file: zeus_def.tex
\newcommand{\ZcoosysB}{%
The ZEUS coordinate system is a right-handed Cartesian system, with the $Z$
axis pointing in the proton beam direction, referred to as the ``forward
direction'', and the $X$ axis pointing left towards the centre of HERA.
The coordinate origin is at the nominal interaction point.\xspace}
\newcommand{\Zpsrap}{%
The pseudorapidity is defined as $\eta=-\ln\left(\tan\frac{\theta}{2}\right)$,
where the polar angle, $\theta$, is measured with respect to the proton beam
direction.\xspace}

\newcommand{\ZcoosysfnBeta}{\footnote{\ZcoosysB\Zpsrap}}

\newcommand{\Zctddesc}[1]{%
Charged particles are tracked in the central tracking detector (CTD)~\citeCTD,
which operates in a magnetic field of $1.43\Tesla$ provided by a thin 
superconducting coil. The CTD consists of 72~cylindrical drift chamber 
layers, organized in 9~superlayers covering the polar-angle#1 region 
\mbox{$15^\circ<\theta<164^\circ$}. The transverse-momentum resolution for
full-length tracks is $\sigma(p_T)/p_T=0.0058p_T\oplus0.0065\oplus0.0014/p_T$,
with $p_T$ in $\Gev$.}

\newcommand{\DO}{{D{\O}}\xspace}

\chardef\usc=95
\chardef\til=126
\catcode`\@=11 
\DeclareRobustCommand\xdotspace{\futurelet\@let@token\@xdotspace}
\def\@xdotspace{%
  \ifx\@let@token.\else
  \ifx\@let@token\bgroup.\else
  \ifx\@let@token\egroup.\else
  \ifx\@let@token\/.\else
  \ifx\@let@token\ .\else
  \ifx\@let@token~.\else
  \ifx\@let@token!.\else
  \ifx\@let@token,.\else
  \ifx\@let@token:.\else
  \ifx\@let@token;.\else
  \ifx\@let@token?.\else
  \ifx\@let@token/.\else
  \ifx\@let@token'.\else
  \ifx\@let@token).\else
  \ifx\@let@token-.\else
  \ifx\@let@token\@xobeysp.\else
  \ifx\@let@token\space.\else
  \ifx\@let@token\@sptoken.\else
   .\space
   \fi\fi\fi\fi\fi\fi\fi\fi\fi\fi\fi\fi\fi\fi\fi\fi\fi\fi}
\catcode`\@=12 

\newcommand{\stru}[2]{%
   \relax\ifmmode\hbox{\vrule height#1 depth#2 width0pt}%
   \else\vrule height#1 depth#2 width0pt\fi}

\newcommand{\Ronum}[1]{\uppercase\expandafter{\romannumeral#1}}
\newcommand{\ronum}[1]{\expandafter{\romannumeral#1}}
\DeclareRobustCommand{\LaTeXZ}{%
  \LaTeX\kern-.05em4\kern-.1em
  {\raisebox{-0.2ex}{$\scriptstyle\text{ZEUS}$}}\xspace}

\DeclareMathAlphabet{\mathbf}{OT1}{cmr}{bx}{sl}
\newcommand{\eVdist}{\kern-0.06667em}

\newcommand{\Gev}{{\text{Ge}\eVdist\text{V\/}}}

\newcommand{\gev}{{\,\text{Ge}\eVdist\text{V\/}}}

\newcommand{\Tesla}{\,\text{T}}

\newcommand{\slashfrac}[2]{%
  \raisebox{0.5ex}{\ensuremath #1}\kern-0.12em/\kern-0.08em
  \raisebox{-.8ex}{\ensuremath #2}}

\newcommand{\sqr}[3]{%
    {\vcenter{\hrule height.#3ex\hbox{\vrule width.#2ex height#1ex
     \kern#1ex\vrule width.#3ex}\hrule height.#2ex}}}

\newcommand{\widebar}[1]{%
   \mkern1.5mu\overline{\mkern-1.5mu#1\mkern-1.mu}\mkern1.mu}
\catcode`\@=11 
\newcommand{\parenbar}{\mathpalette\p@renb@r}
\def\p@renb@r#1#2{\vbox{%
  \ifx#1\scriptscriptstyle \dimen@.7em\dimen@ii.2em\else
  \ifx#1\scriptstyle \dimen@.8em\dimen@ii.25em\else
  \dimen@1em\dimen@ii.4em\fi\fi \offinterlineskip
  \ialign{\hfill##\hfill\cr
    \vbox{\hrule width\dimen@ii}\cr
    \noalign{\vskip-.3ex}%
    \hbox to\dimen@{$\mathchar300\hfil\mathchar301$}\cr
    \noalign{\vskip-.3ex}%
    $#1#2$\cr}}}
\catcode`\@=12 

\newcommand{\pbar}{\widebar{p}}

\newcommand{\IP}{{\rm I$\kern-0.01667em$P}\xspace}

\mathchardef\qsm=63
\mathchardef\pls=43
\mathchardef\mns=512
\mathchardef\plm=518
\mathchardef\eql=61
\mathchardef\smallleft=300
\mathchardef\smallright=301
\mathchardef\les=316
\mathchardef\gre=318
\mathchardef\leq=532
\mathchardef\grq=533

\catcode`\@=11 
\newcounter{pict@width}
\newcounter{pict@height}
\newlength{\pict@scale}
\newcommand{\psfigadd}[4]{%
\setcounter{pict@width}{1*\ratio{#2+\pict@scale/2}{\pict@scale}}
\setcounter{pict@height}{1*\ratio{#3+\pict@scale/2}{\pict@scale}}
\setlength{\unitlength}{\pict@scale}
\hbox to #2{\hspace{-\fill}\begin{picture}(\thepict@width,\thepict@height)
\put(0,0){\psfig{figure=#1,width=#2,height=#3,clip=}}
\SetScale{0.283466457}
\SetWidth{1.763889}
{#4}
\end{picture}}
}
\newcounter{pict@widthfst}
\newcounter{pict@widthscd}
\newcounter{pict@widthtot}
\newcommand{\psfigaddtwo}[7]{%
\setcounter{pict@widthfst}{1*\ratio{#2+\pict@scale/2}{\pict@scale}}
\setcounter{pict@widthscd}{1*\ratio{#2+#4+\pict@scale/2}{\pict@scale}}
\setcounter{pict@widthtot}{1*\ratio{#2+#4+#6+\pict@scale/2}{\pict@scale}}
\setcounter{pict@height}{1*\ratio{#3+\pict@scale/2}{\pict@scale}}
\setlength{\unitlength}{\pict@scale}
\hbox{\hspace{-\fill}\begin{picture}(\thepict@widthtot,\thepict@height)
\put(0,0){\psfig{figure=#1,width=#2,height=#3,clip=}}
\put(\thepict@widthscd,0){\psfig{figure=#5,width=#6,height=#3,clip=}}
\SetScale{0.283466457}
\SetWidth{1.763889}
{#7}
\end{picture}}
}
\newcommand{\psfigror}[4]{%
\setcounter{pict@width}{1*\ratio{#2+\pict@scale/2}{\pict@scale}}
\setcounter{pict@height}{1*\ratio{#3+\pict@scale/2}{\pict@scale}}
\setlength{\unitlength}{\pict@scale}
\hbox{\begin{picture}(\thepict@width,\thepict@height)
\put(0,\thepict@height){\psfig{figure=#1,width=#3,height=#2,clip=,angle=270}}
\SetScale{0.283466457}
\SetWidth{1.763889}
{#4}
\end{picture}}
}
\newcommand{\psfigrol}[4]{%
\setcounter{pict@width}{1*\ratio{#2+\pict@scale/2}{\pict@scale}}
\setcounter{pict@height}{1*\ratio{#3+\pict@scale/2}{\pict@scale}}
\setlength{\unitlength}{\pict@scale}
\hbox{\begin{picture}(\thepict@width,\thepict@height)
\put(0,0){\psfig{figure=#1,width=#3,height=#2,clip=,angle=90}}
\SetScale{0.283466457}
\SetWidth{1.763889}
{#4}
\end{picture}}
}
\catcode`\@=12 
\newlength\listtextwidth

\catcode`\@=11 
\newlength{\@tabfninsert}
\newlength{\@tabfnwidth}
\newcommand{\tabfootnote}[2]{%
  \setlength{\@tabfninsert}{0.8em}
  \setlength{\@tabfnwidth}{\textwidth}
  \addtolength{\@tabfnwidth}{-\@tabfninsert}
  \addtolength{\@tabfnwidth}{-0.4em}
  \noindent\makebox[\@tabfninsert][r]{\footnotesize$^{#1}$\hfil}\hfill%
  \parbox[t]{\@tabfnwidth}{\footnotesize #2\hfill}}
\catcode`\@=12 

%% file: DESY-06-166.bbl
\begin{mcbibliography}{10}

\bibitem{np:b412:225}
S.~Frixione \etal,
\newblock Nucl.\ Phys.{} {\bf B~412},~225~(1994)\relax
\relax
\bibitem{np:b454:3-24}
S. Frixione, P. Nason and G. Ridolfi,
\newblock Nucl.\ Phys.{} {\bf B 454}~(1995)\relax
\relax
\bibitem{frixione3}
M. Cacciari, S. Frixione and P. Nason,
\newblock JHEP{} {\bf 0103},~006~(2001)\relax
\relax
\bibitem{np:b452:109}
B.W.~Harris and J.~Smith,
\newblock Nucl.\ Phys.{} {\bf B~452},~109~(1995)\relax
\relax
\bibitem{pl:b353:535}
B.~W.~Harris and J.~Smith,
\newblock Phys.\ Lett.{} {\bf B~353},~535~(1995).
\newblock Erratum-ibid {\bf B~359} (1995) 423\relax
\relax
\bibitem{pr:d57:2806}
B.W.~Harris and J.~Smith,
\newblock Phys.\ Rev.{} {\bf D~57},~2806~(1998)\relax
\relax
\bibitem{beautyUA10}
UA1 \coll, C.~Albajar \etal,
\newblock Phys. Lett.{} {\bf B 186},~237~(1987)\relax
\relax
\bibitem{pl:b256:121-128}
UA1 \coll, C.~Albajar \etal,
\newblock Phys.\ Lett.{} {\bf B 256},~121~(1991).
\newblock Erratum in Phys.~Lett.~{\bf{B~262}}, 497 (1991)\relax
\relax
\bibitem{zfp:c61:41-52}
UA1 \coll, C.~Albajar \etal,
\newblock Z.\ Phys.{} {\bf C 61},~41~(1994)\relax
\relax
\bibitem{beautyCDF1}
CDF \coll, F. Abe et al.,
\newblock Phys. Rev. Lett.{} {\bf 71},~500~(1993)\relax
\relax
\bibitem{beautyCDF2}
CDF \coll, F. Abe et al.,
\newblock Phys. Rev. Lett.{} {\bf 71},~2396~(1993)\relax
\relax
\bibitem{beautyCDF3}
CDF \coll, F. Abe et al.,
\newblock Phys. Rev. Lett.{} {\bf 75},~1451~(1995)\relax
\relax
\bibitem{beautyCDF4}
CDF \coll, F. Abe et al.,
\newblock Phys. Rev.{} {\bf D 53},~1051~(1996)\relax
\relax
\bibitem{beautyCDF5}
CDF \coll, D. Acosta et al.,
\newblock Phys. Rev.{} {\bf D~65},~052005~(2002)\relax
\relax
\bibitem{beautyCDF6_pap}
CDF \coll, D. Acosta et al.,
\newblock Phys. Rev.{} {\bf D~71},~032001~(2005)\relax
\relax
\bibitem{beautyD00}
\DO \coll, S. Abachi et al.,
\newblock Phys. Rev. Lett.{} {\bf 74},~3548~(1995)\relax
\relax
\bibitem{beautyD01}
\DO \coll, B. Abbott et al.,
\newblock Phys. Lett.{} {\bf B 487},~264~(2000)\relax
\relax
\bibitem{beautyD02}
\DO \coll, B. Abbott et al.,
\newblock Phys. Rev. Lett.{} {\bf 84},~5478~(2000)\relax
\relax
\bibitem{beautyD03}
\DO \coll, B. Abbott et al.,
\newblock Phys. Lett.{} {\bf 85},~5068~(2000)\relax
\relax
\bibitem{beautyLEP0}
L3 \coll, M.~Acciarri \etal,
\newblock Phys. Lett.{} {\bf B~503},~10~(2000)\relax
\relax
\bibitem{beautyLEP1}
L3 \coll, P.~Achard \etal,
\newblock Phys. Lett.{} {\bf B~619},~71~(2005)\relax
\relax
\bibitem{WA78}
WA78 \coll, M. Catanesi et al.,
\newblock Phys. Lett.{} {\bf B 202},~453~(1988)\relax
\relax
\bibitem{E706}
E672/E706 \coll, R. Jesik et al.,
\newblock Phys. Rev. Lett.{} {\bf 74},~495~(1995)\relax
\relax
\bibitem{E771}
E771 \coll, T. Alexopoulos et al.,
\newblock Phys. Rev. Lett.{} {\bf 82},~41~(1999)\relax
\relax
\bibitem{E789}
D.M. Jansen et al.,
\newblock Phys. Rev. Lett.{} {\bf 74},~3118~(1995)\relax
\relax
\bibitem{HERAB}
HERA--B \coll, I. Abt et al.,
\newblock Eur. Phys. J.{} {\bf C 26},~345~(2003)\relax
\relax
\bibitem{pl:b467:156}
H1 \coll, C.~Adloff \etal,
\newblock Phys.\ Lett.{} {\bf B~467},~156~(1999)\relax
\relax
\bibitem{epj:c18:625}
ZEUS \coll, J.~Breitweg \etal,
\newblock Eur.\ Phys.\ J.{} {\bf C~18},~625~(2001)\relax
\relax
\bibitem{epj:c40:349}
H1 \coll, A.~Aktas \etal,
\newblock Eur.\ Phys.\ J.{} {\bf C~40},~349~(2005)\relax
\relax
\bibitem{epj:c41:453}
H1 \coll, A.~Aktas \etal,
\newblock Eur.\ Phys.\ J.{} {\bf C~41},~453~(2005)\relax
\relax
\bibitem{Aktas:2005iw}
H1 \coll, A.~Aktas \etal,
\newblock Eur. Phys. J.{} {\bf C~45},~23~(2006)\relax
\relax
\bibitem{H1phjets}
H1 \coll, A. Aktas et al.,
\newblock Eur. Phys. J.{} {\bf C~47},~597~(2006)\relax
\relax
\bibitem{pr:d70:012008}
ZEUS \coll, S.~Chekanov \etal,
\newblock Phys.\ Rev.{} {\bf D~70},~12008~(2004).
\newblock Erratum accepted by Phys.~Rev.~D, hep-ex/0312057\relax
\relax
\bibitem{pl:b599:173}
ZEUS \coll, S.~Chekanov \etal,
\newblock Phys.\ Lett.{} {\bf B~599},~173~(2004)\relax
\relax
\bibitem{pl:b621:56-71}
H1 \coll, A.~Aktas \etal,
\newblock Phys.\ Lett.{} {\bf B~621},~56~(2005)\relax
\relax
\bibitem{zeus:1993:bluebook}
ZEUS \coll, U.~Holm~(ed.),
\newblock {\em The {ZEUS} Detector}.
\newblock Status Report (unpublished), DESY (1993),
\newblock available on
  \texttt{http://www-zeus.desy.de/bluebook/bluebook.html}\relax
\relax
\bibitem{nim:a279:290}
N.~Harnew \etal,
\newblock Nucl.\ Inst.\ Meth.{} {\bf A~279},~290~(1989)\relax
\relax
\bibitem{npps:b32:181}
B.~Foster \etal,
\newblock Nucl.\ Phys.\ Proc.\ Suppl.{} {\bf B~32},~181~(1993)\relax
\relax
\bibitem{nim:a338:254}
B.~Foster \etal,
\newblock Nucl.\ Inst.\ Meth.{} {\bf A~338},~254~(1994)\relax
\relax
\bibitem{nim:a309:77}
M.~Derrick \etal,
\newblock Nucl.\ Inst.\ Meth.{} {\bf A~309},~77~(1991)\relax
\relax
\bibitem{nim:a309:101}
A.~Andresen \etal,
\newblock Nucl.\ Inst.\ Meth.{} {\bf A~309},~101~(1991)\relax
\relax
\bibitem{nim:a321:356}
A.~Caldwell \etal,
\newblock Nucl.\ Inst.\ Meth.{} {\bf A~321},~356~(1992)\relax
\relax
\bibitem{nim:a336:23}
A.~Bernstein \etal,
\newblock Nucl.\ Inst.\ Meth.{} {\bf A~336},~23~(1993)\relax
\relax
\bibitem{nim:a401:63}
A.~Bamberger \etal,
\newblock Nucl.\ Inst.\ Meth.{} {\bf A~401},~63~(1997)\relax
\relax
\bibitem{epj:c21:443}
ZEUS \coll, S.~Chekanov \etal,
\newblock Eur.\ Phys.\ J.{} {\bf C~21},~443~(2001)\relax
\relax
\bibitem{nim:a333:342}
G.~Abbiendi \etal,
\newblock Nucl.\ Inst.\ Meth.{} {\bf A~333},~342~(1993)\relax
\relax
\bibitem{Desy-92-066}
J.~Andruszk\'ow \etal,
\newblock Preprint \mbox{DESY-92-066}, DESY, 1992\relax
\relax
\bibitem{zfp:c63:391}
ZEUS \coll, M.~Derrick \etal,
\newblock Z.\ Phys.{} {\bf C~63},~391~(1994)\relax
\relax
\bibitem{acpp:b32:2025}
J.~Andruszk\'ow \etal,
\newblock Acta Phys.\ Pol.{} {\bf B~32},~2025~(2001)\relax
\relax
\bibitem{np:b729:492-525}
ZEUS \coll, S.~Chekanov \etal,
\newblock Nucl.\ Phys.{} {\bf B~729},~492~(2005)\relax
\relax
\bibitem{proc:epfacility:1979:391}
F.~Jacquet and A.~Blondel,
\newblock {\em Proceedings of the Study for an $ep$ Facility for {Europe}},
  U.~Amaldi~(ed.), p.~391.
\newblock Hamburg, Germany (1979).
\newblock Also in preprint \mbox{DESY 79/48}\relax
\relax
\bibitem{cpc:82:74}
T.~Sj\"ostrand,
\newblock Comp.\ Phys.\ Comm.{} {\bf 82},~74~(1994)\relax
\relax
\bibitem{cpc:86:147}
H.~Jung,
\newblock Comp.\ Phys.\ Comm.{} {\bf 86},~147~(1995)\relax
\relax
\bibitem{cpc:67:465}
G.~Marchesini \etal,
\newblock Comp.\ Phys.\ Comm.{} {\bf 67},~465~(1992)\relax
\relax
\bibitem{thesis:longhin:2004}
A. Longhin, Ph.D. Thesis, Report \mbox{DESY-THESIS-2004-050}, Universit\`a di
  Padova and INFN, 2004\relax
\relax
\bibitem{GEANT}
R.~Brun et al.,
\newblock {\em {\sc geant3}},
\newblock Technical Report CERN-DD/EE/84-1, CERN, 1987\relax
\relax
\bibitem{bmassless1}
J. Binnewies, B.A. Kniehl and G. Kramer,
\newblock Z. Phys.{} {\bf C~76},~677~(1997)\relax
\relax
\bibitem{bmassless2}
B.A. Kniehl, G. Kramer and M. Spira,
\newblock Z. Phys.{} {\bf C~76},~689~(1997)\relax
\relax
\bibitem{bmassless3}
J. Binnewies, B.A. Kniehl and G. Kramer,
\newblock Phys. Rev.{} {\bf D~58},~014014~(1998)\relax
\relax
\bibitem{bkt1}
A. Lipatov and N. Zotov,
\newblock Preprint \mbox{hep-ph/0601240}, 2006\relax
\relax
\bibitem{bkt2}
A. Lipatov and N. Zotov,
\newblock Preprint \mbox{hep-ph/0603017}, 2006\relax
\relax
\bibitem{jung1}
H.~Jung,
\newblock Phys. Rev.{} {\bf D 65},~034015~(2002)\relax
\relax
\bibitem{jung2}
H.~Jung,
\newblock J. Phys.{} {\bf G 28},~971~(2002)\relax
\relax
\bibitem{epj:c12:375}
CTEQ \coll, H.L.~Lai \etal,
\newblock Eur.\ Phys.\ J.{} {\bf C~12},~375~(2000)\relax
\relax
\bibitem{pr:d46:1973-1979}
M.~Gl\"uck, E.~Reya, and A.~Vogt,
\newblock Phys.\ Rev.{} {\bf D~46},~1973~(1992)\relax
\relax
\bibitem{LizDIS}
Adriana Elizabeth Nuncio-Quiroz, ZEUS \coll,
\newblock {\em \rm{talk at XIV International Workshop on Deep Inelastic
  Scattering, Tsukuba, Japan, 20-24 April 2006; to appear in the
  proceedings}}\relax
\relax
\bibitem{hep-ph-0108264}
T.~Sj\"ostrand, L.~L\"onnblad, and S.~Mrenna,
\newblock Preprint \mbox{hep-ph/0108264}, 2001\relax
\relax
\bibitem{jhep:06:029}
S.~Frixione and B.R.~Webber,
\newblock JHEP{} {\bf 06},~029~(2002)\relax
\relax
\bibitem{jhep:08:007}
S.~Frixione, P.~Nason and B.R.~Webber,
\newblock JHEP{} {\bf 08},~007~(2003)\relax
\relax
\bibitem{pr:d27:105}
C.~Peterson \etal,
\newblock Phys.\ Rev.{} {\bf D~27},~105~(1983)\relax
\relax
\bibitem{zfp:c20:317}
B.~Andersson, G.~Gustafson and B.~S\"oderberg,
\newblock Z.\ Phys.{} {\bf C~20},~317~(1983)\relax
\relax
\bibitem{np:b238:492}
B.R.~Webber,
\newblock Nucl.\ Phys.{} {\bf B~238},~492~(1984)\relax
\relax
\bibitem{cpc:39:347}
T.~Sj\"ostrand,
\newblock Comp.\ Phys.\ Comm.{} {\bf 39},~347~(1986)\relax
\relax
\bibitem{pl:b592:1}
Particle Data Group, S.~Eidelman \etal,
\newblock Phys.\ Lett.{} {\bf B~592},~1~(2004)\relax
\relax
\bibitem{zfp:88:612}
C.F.~von Weizs\"acker,
\newblock Z.\ Phys.{} {\bf 88},~612~(1934)\relax
\relax
\bibitem{pr:45:729}
E.J.~Williams,
\newblock Phys.\ Rev.{} {\bf 45},~729~(1934)\relax
\relax
\bibitem{pl:b319:339-345}
S.~Frixione \etal,
\newblock Phys.\ Lett.{} {\bf B~319},~339~(1993)\relax
\relax
\bibitem{pr:d69:012004}
ZEUS \coll, S.~Chekanov \etal,
\newblock Phys.\ Rev.{} {\bf D~69},~012004~(2004)\relax
\relax
\bibitem{pl:b590:143}
ZEUS \coll, S.~Chekanov \etal,
\newblock Phys.\ Lett.{} {\bf B~590},~143~(2004)\relax
\relax
\end{mcbibliography}
